\documentclass[preprint,12pt]{elsarticle}
\usepackage{amssymb}
\begin{document}
\begin{frontmatter}
\title{An Analytical Framework for Imposition of a Rigid Immersed Surface on the  Incompressible Navier--Stokes Equations}
\author{Fereidoun Sabetghadam}
\address{Mechanical and Aerospace Engineering Department, Science and Research Branch, Azad University (IAU), Tehran, Iran}
\begin{abstract}
The incompressible Navier--Stokes equations are re-formulated to involve an arbitrary time dilation; and in this manner, the modified Navier--Stokes equations are obtained which have some penalization terms in the right hand side. Then, the solid rigid bodies are modeled as the regions where time is dilated infinitely. The physical and mathematical properties of the modified equations and the penalization terms are investigated, and it is shown that the modified equations satisfy the no-slip, no-diffusion, no-advection, and no-pressure coupling conditions. The modified equations can be used in exact imposition of the solid rigid bodies on the incompressible Navier--Stokes equations. To show the capability of the modified equations, three classical exact solutions of the Navier--Stokes equations, that is, the Stokes first problem, the plane stagnation point flow, and the stokes flow over a sphere are re-solved exactly, this time in the presence of a solid region. 
\end{abstract}
\end{frontmatter}
\section{Introduction}\label{s1}
Efficient imposition of the solid boundaries on the incompressible Navier--Stokes equations has been one of the longstanding challenges of the theoretical as well as computational fluid dynamics. The effects of the solid boundaries on the fluid flow are commonly modeled by the no-slip condition, which means the velocities of the fluid and solid are equal at the solid boundaries.

In spite of the extensive body of literature on the implementation of the no-slip condition, both for the regular boundaries [3, 5, 2], and the immersed boundaries [6, 7], it seems that exact implementation of this condition so that the velocity vector remains solenoidal all over the solution domain is still an unsolved problem. \\
The present article, suggests modification of the incompressible Navier--Stokes equations so that the solid rigid bodies can be implemented exactly. By the term rigid we mean a material of infinite elastic modulus. This is a simplified model which may be used in many fluid--solid problems, at least approximately. In these materials, by their nature, the distribution of stresses is not a continuous function. In particular, the distribution of stresses in the body is not depended on the stresses on the boundaries. It means the stresses on the boundary of a rigid body, in a fluid--solid system, do not affect the stresses inside the rigid body. This is a mutual effect, that is, the stresses in the rigid body do not affect the stresses in the fluid as well. In fact, the stresses field in the fluid--solid system are discontinuous at the solid surface. Therefore, the rigid body surface is a singularity surface in that the information do not exchange between its two sides.\\ 
Such singularity surfaces are well-known in the general relativity and cosmology. In fact, the black holes have been in studying for several decades as the singularities in the space--time structure [4]. The black hole surface (the event horizon, as it is called in the general relativity) is assumed to be a singularity surface that no information can be exchanged between its two sides. For the black holes, the singularity is due to infinite curvature in the space--time manifold which, in turn, is a result of extremely strong gravity of the black hole, which causes an infinite time dilation.\\  
Now, in a fluid--solid system, if we assume that vanishing of the velocities in the solid body is a result of stopping the time (i.e. infinite time dilation), not the result of properties of the material, then many similarities can be observed between the presence of a black hole in the space--time and the presence of a solid rigid body in an incompressible flow (except for the gravity of coarse). This is the key assumption we made in our study. \\
In the approach we suggest in the present article, the fluid--solid system is modeled by a fluid flow field having two different time dilations. The fluid region has no time dilation, while in the solid region the time dilation goes to infinity. As it will be seen, the presence of the solid body will fully sense by the fluid flow in this way. \\
In the sequel, at first, we re-formulate the incompressible Navier--Stokes equations to involve an arbitrary time dilation. Then, suitable time dilation for imposition of the rigid solid bodies is introduced in the equations and the physical and mathematical characteristics of the resulting modified Naver--Stokes equations are discussed. Finally, the capability of the modified equations in imposition of the solid bodies is demonstrated via solution of three classical exact solutions of the Navier--Stokes equations.
\section{Modifying the incompressible Navier--Stokes equations}
The modified Navier--Stokes equations are derived in this section. Beginning from the classical incompressible Navier--Stokes equations, an arbitrary time dilation is imposed on the equations and then a particular time dilation is proposed so that the rigid bodies can be modeled. The physical and mathematical properties of the modified equations are discussed in the end. 
\begin{figure}
\centering
\includegraphics[width=0.5\textwidth]{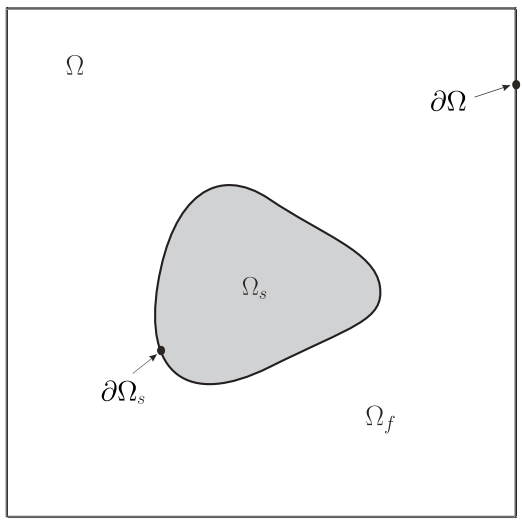}
\caption{A typical fluid-solid (rigid) body problem. The solid body $\Omega_s$ with its boundary $\partial \Omega_s$ is placed in the flow of an incompressible Newtonian fluid occupies the flow domain $\Omega_f$. The fluid flow is formulated and solved in a regular domain $\Omega$ with its boundary $\partial \Omega$.} \label{fig0}
\end{figure}
\subsection{Preliminaries and basic definitions}
According to Fig. \ref{fig0}, consider a regular domain $\Omega\in {\Bbb R}^d$ where $d=2$ or $3$ and its boundary $\partial \Omega$, containing a rigid solid obstacle $\Omega_s$ with its boundary $\partial \Omega_s$. Moreover, assume that the fluid domain $\Omega_f=\Omega \setminus \bar{\Omega}_s$ (where $\bar{\Omega}_s=\Omega_s \cup \partial \Omega_s$),  is occupied by an incompressible Newtonian fluid that its dynamics is governed by the incompressible Navier--Stokes equations  
\begin{equation}\label{e0}
\left\{
\begin{array}{rl}
&\partial_t {\tilde{\bf u}}+(\tilde{{\bf u}}^T\cdot \nabla){\tilde{\bf u}}=-\frac{1}{\tilde{\varrho}}\nabla \tilde{p}+\tilde{\nu}\nabla^2{\tilde{\bf u}} \quad \quad {\rm in}\quad \Omega_f \times {\Bbb R}^+,\\
&\nabla \cdot {\tilde{\bf u}}=0~ \quad \quad \quad  \quad \quad \quad \quad \quad \quad \quad \quad ~ ~~~ {\rm in}~\quad \Omega_f ,
\end{array} \right.
\end{equation}
in which ${\tilde{\bf u}}=[\tilde{u}_1\cdots \tilde{u}_d]^T$ is the velocity vector, $\tilde{p}$ is the pressure, $\tilde{\varrho}$ and $\tilde{\nu}$ are the fluid density and the kinematic viscosity. The initial and boundary conditions for the above system are: 
\begin{eqnarray}
&&{\rm{IC:~~~}}{\tilde{\bf u}}({\bf x},0)={\tilde{\bf u}}_0({\bf x}) \quad \quad {\rm in}\quad \Omega_f , \label{e011} \\&& \nonumber \\
&&{\rm BCs:~~}{\tilde{\bf u}}(\partial\Omega,t)={\tilde{\bf u}}_{\partial\Omega}(t),\label{e01}\\
&&~~~~~~~~~~  {\tilde{\bf u}}(\partial{\Omega_s},t)=0,\label{e02}
\end{eqnarray}
where we assume that the initial velocity is solenoidal $\nabla\cdot{\tilde{\bf u}}_0(\Omega_f)=0 $; and the initial and boundary data are compatible. 

For the future usage, system (\ref{e0}) with the initial/boundary conditions (\ref{e011})--(\ref{e02}) is called ${\Bbb S}_0$. At this point we assume that system ${\Bbb S}_0$ has a unique solution, which means there are sufficiently smooth velocity vector filed ${\tilde{\bf u}}({\bf x},t)$ and pressure field $\tilde{p}({\bf x},t)$ that satisfy equations (\ref{e0}) and the boundary conditions (\ref{e011})--(\ref{e02}). 
\subsection{The Incompressible Navier-Stokes equations in the presence of time dilation}
Now, it is desired to re-formulate the incompressible Navier--Stokes (without rigid obstacles) in the presence of an arbitrary local time dilation. The formulation is provided in the regular domain $\Omega$ (see Fig. \ref{fig0}).

In the regular domain $\Omega\in {\Bbb R}^d$ and its boundary $\partial \Omega$, occupied by a Newtonian fluid, we consider two distinct inertial space--time coordinates; that is, a reference space--time coordinate $({\bf x},t)$ and a local space--time coordinate $({\bf x}^*,t^*)$. Merely the time dilation is desired (without any changes in the space or other). Therefore, we assume ${\bf x}^*={\bf x}$ while $t^*=t/\lambda$ where the time dilation factor $\lambda=\lambda({\bf x},t)$ is given at each reference time instant $t$. The fluid flow is observed by both local and reference observers; however, the time dilation is recognizable only by the reference observer.\\ 

Since the physical laws are invariant in all inertial coordinates [4], the Navier--Stokes equations (\ref{e0}) are valid in the local coordinate $({\bf x}^*,t^*)\equiv({\bf x},t^*)$, that is 
\begin{equation}\label{e1}
\left\{
\begin{array}{rl}
&\partial_t {\bf u}^*+({{\bf u}^*}^T\cdot \nabla){\bf u}^*=-\frac{1}{\varrho^*}\nabla p^*+\nu^*\nabla^2{\bf u}^* \quad \quad {\rm in}\quad \Omega \times {\Bbb R}^+,\\
&\nabla \cdot {\bf u}^*=0 \quad \quad \quad \quad  \quad \quad \quad \quad \quad \quad \quad \quad \quad \quad ~~~ {\rm in}~\quad \Omega ,
\end{array} \right.
\end{equation}
with the initial/boundary conditions
\begin{eqnarray}
&&{\rm{IC:~~~}}{\bf u}^*({\bf x},0)={\bf u}^*_0({\bf x}), \quad \quad {\rm in}\quad \Omega , \label{e0111} \\ 
&&{\rm BCs:~}{\bf u}^*(\partial\Omega,t)={\bf u}^*_{\partial\Omega}(t).\label{e01111}
\end{eqnarray}
In these equations $(\cdot)^*$ are quantities that are observed from the local coordinate. Particularly, note that $\varrho^*$ and $\nu^*$ are the molecular density and viscosity, which assumed to be constant. Moreover, we assume that the initial velocity is solenoidal (i.e. $\nabla\cdot {\bf u}_0^*=0$), and the initial and boundary data are compatible.\\
In the next sections, system (\ref{e1}) with the initial/boundary conditions (\ref{e0111})--(\ref{e01111}) will be referred as ${\Bbb S}_1$.\\

\noindent
{\bf Remark 1:} It should be noted that the only boundary is $\partial \Omega$ (and therefore, the only required boundary condition is on $\partial \Omega$); and  the no-slip condition is eliminated for now (In fact it will be imposed by the forcing terms in the future.)  \\ 

Now we can obtain the above equations in the viewpoint of the reference space--time coordinate $({\bf x},t)$. To this end:
\begin{itemize}
\item[{\bf i)}]  {\bf The velocities}: One can write
\begin{equation}
{\bf u}^*=\frac{d{\bf x}^*}{dt^*}=\frac{d{\bf x}}{dt^*}=\lambda \frac{d{\bf x}}{dt}=\lambda {\bf u},
\label{u_p}
\end{equation}
where ${\bf u}$ is observed from the reference coordinate.
\item[{\bf ii)}] {\bf The pressure and the density}: In the incompressible Navier--Stokes equations the pressure is not a thermodynamics quantity. Therefore, the relation between the local and reference pressures cannot be determined uniquely by merely determining the time dilation.\\
In general, one can say
\begin{equation}
{\rm Fluid~particles~acceleration~ }=\frac{Du^*}{Dt^*}\sim -\frac{1}{\varrho^*} \frac{d p^*}{dx},
\end{equation}  
where $\frac{D}{Dt}$ is the material derivative and $x$ is a space coordinate. Therefore
\begin{equation}
 -\frac{1}{\varrho^*} \frac{d p^*}{dx}\sim \lambda^2\frac{Du}{Dt}.
\end{equation}
This is the only restriction that the incompressible Navier--Stokes equations is enforced. Therefore, there are two degrees of freedom (that is, the relations between the local and reference density and pressure could be determined concurrently). In fact, any $p^*=\lambda^m p$ and $\varrho^*=\lambda^n\varrho$ that $m-n=2$  results in the momentum equations that their two sides are consistent; and produces a pressure jump in crossing the solid boundary (which physically is legitimated).  

However, in order to have  a model that is as simple as possible, we choose $n=0$, which means $m=2$, that is 
\begin{eqnarray}
&& p^*=\lambda^2 p,\\
&& \varrho^*=\varrho.
\end{eqnarray}
\item[{\bf iii)}] {\bf The viscosity}: The Reynolds number must be an invariant quantity (it is not physically meaningful looking at a low Reynolds number flow and observing turbulence.) Therefore:
\[
{\rm Re}=\frac{UL}{\nu}={\rm Re}^*=\frac{U^* L}{\nu^*}=\frac{\lambda UL}{\nu^*} 
\]
which means 
\begin{equation}
\nu^*=\lambda \nu.
\end{equation}
In fact, the viscosity observed by the reference observer (i.e. $\nu$), is a function of $\lambda$. On the other hand, substitution from $\varrho$ results in
\begin{equation}
\mu^*=\varrho^*\nu^*=\varrho(\lambda\nu)=\lambda\mu.
\end{equation}
\end{itemize}
Using the above assumptions, one can find the Navier--Stokes equations in the viewpoint of the reference coordinate $({\bf x},t)$; by substituting $t^*$, ${\bf u}^*$, $p^*$, $\varrho^*$ and $\nu^*$ in Eqns. (\ref{e1}), and using the following vector identities:
\begin{eqnarray}
&&\nabla \cdot (\lambda {\bf u})=\lambda \nabla {\bf u}+({\bf u}^T\cdot \nabla) \lambda, \nonumber \\
&&[(\lambda {\bf u})^T\cdot \nabla](\lambda {\bf u})=\lambda^2({\bf u}^T\cdot \nabla){\bf u}+\lambda{\bf u}({\bf u}^T\cdot \nabla\lambda), \label{identity}\\
&&\nabla^2(\lambda{\bf u})=\lambda \nabla^2{\bf u}+{\bf u}\nabla^2\lambda +2\nabla {\bf u}\cdot \nabla \lambda. \nonumber
\end{eqnarray}
In fact, it can easily be verified that using the above relations, the Navier--Stokes equations observing from the reference space--time coordinate are:
\begin{equation}\label{e7}
\left\{
\begin{array}{rl}
&\partial_t {\bf u}+({\bf u}^T\cdot \nabla){\bf u}=-\frac{1}{\varrho}\nabla p+\nu\nabla^2{\bf u}+{\bf F}^{\lambda},\\
&\nabla \cdot {\bf u}={\bf G}^{\lambda}.
\end{array} \right.
\end{equation} 
in which 
\begin{equation}\label{e8}
\left\{
\begin{array}{rl}
&{\bf F}^{\lambda}=\frac{1}{\lambda}[\nu {\bf u}\nabla^2\lambda+2\nu\nabla {\bf u}\cdot \nabla \lambda-{\bf u}({\bf u}^T\cdot \nabla)\lambda-2\frac{p}{\varrho}\nabla \lambda],\\
&{\bf G}^{\lambda}=-\frac{1}{\lambda}({\bf u}^T\cdot \nabla)\lambda.
\end{array} \right.
\end{equation}
\noindent 
This system of equations will be called the modified Navier--Stokes equations, and will be used in imposition of rigid immersed surfaces on the incompressible Navier--Stokes equations.\\

\noindent 
With regard to these equations the following points are noticeable:
\begin{itemize}
\item[{\bf (1)}] Both the momentum and continuity equation are modified. In fact, the difference between the modified Navier--Stokes equations (\ref{e7})--(\ref{e8}) and the original Navier--Stokes equations (\ref{e0}) is the forcing terms ${\bf F}^{\lambda}$ and ${\bf G}^{\lambda}$.
\item[{\bf (2)}] The support of both ${\bf F}^{\lambda}$ and ${\bf G}^{\lambda}$ are the support of $\nabla \lambda$. In fact, for any constant time dilation factor $\lambda({\bf x})=\lambda$, both ${\bf F}^{\lambda}$ and ${\bf G}^{\lambda}$ are vanished, and the original Navier--Stokes equations are retrieved. In particular this is in agreement with the general relativity. 
\item[{\bf (3)}] The forcing terms ${\bf F}^\lambda$ and ${\bf G}^\lambda$ were obtained for a general $\lambda$; however, as it will be shown in $\S$ \ref{continuity}, when we impose the rigid surface, finally 
\begin{equation}
{\rm{\bf G}^\lambda}=0,
\label{G0}
\end{equation}
and the classical continuity equation is retrieved.
\item[{\bf (4)}] The boundary and initial conditions for system (\ref{e7})--(\ref{e8}) can be obtained directly by substitution of Eq. (\ref{u_p}) in Eqns. (\ref{e0111}) and (\ref{e01111}). In this manner
\begin{eqnarray}
&&{\rm{IC:~~~}}{\bf u}({\bf x},0)={\bf u}_0({\bf x})=\frac{{\bf u}^*_0}{\lambda} \quad \quad {\rm in}\quad \Omega , \label{e222} \\ 
&&{\rm BCs:~}{\bf u}(\partial\Omega,t)=\frac{{\bf u}^*_{\partial\Omega}(t)}{\lambda}.\label{e22}
\end{eqnarray} 
Moreover, we assume that the initial velocity satisfies the modified continuity equation
\begin{equation}
\nabla\cdot{\bf u}_0=-\frac{1}{\lambda}({\bf u}_0^T\cdot \nabla)\lambda
\end{equation}
and the initial and boundary data are compatible.
\end{itemize}
For the future usage, system (\ref{e7})--(\ref{e8}) with the initial/boundary conditions (\ref{e222})--(\ref{e22}) is called ${\Bbb S}_2$.\\

Now these modified equations can be used in imposing solid bodies in the fluid flow.
\begin{figure}
\centering
\includegraphics[width=0.55\textwidth]{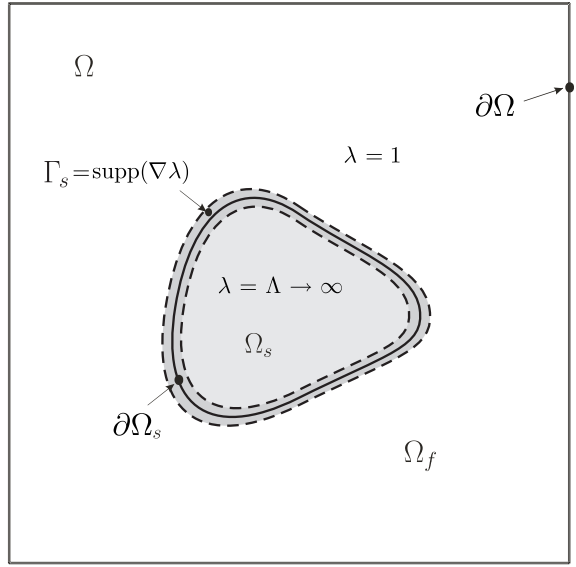}
\caption{Imposition of solid body by time dilation. In the solid body ${\Omega}_s$ the time dilation gets a large positive number $\lambda(\Omega_s)=\Lambda\rightarrow\infty$, while $\lambda(\Omega_f)=1$. In the interface region $\Gamma_s={\mathrm{ supp}}(\nabla \lambda)$, we have $1\leq \lambda \leq \Lambda$.} \label{fig1}
\end{figure}
\subsection{Imposition of a rigid body using the time dilation} 
Presence of a solid body can be implemented to the modified equations  (\ref{e7})--(\ref{e8}) by letting $\lambda\rightarrow+\infty$ in the solid region while leaving $\lambda=1$ in the fluid region.

More precisely, according to Fig. \ref{fig1}, we assume that $\Omega$ consists of two regions; the fluid region $\Omega_f$, and the solid region $\Omega_s$ (i.e., $\Omega=\Omega_f \cup \Omega_s$). Now, we associate a constant positive large number to the time dilation factor in $\Omega_s$, that is $\lambda(\Omega_s)=\Lambda\rightarrow+\infty$; while leave $\lambda(\Omega_f)=1$. Naturally, there is an interface region in the vicinity of $\partial\Omega_s$ that $1 \leq\lambda\leq \Lambda$ (in Fig. \ref{fig1} it is shown by $\Gamma_s={\rm supp}(\nabla \lambda)$ ). 

The sharpness of $\lambda$ in the interface region affects the convergence rate of the solutions to the solution of system ${\Bbb S}_0$, which is our goal. In general, we define the time dilation factor $\lambda$ as  
\begin{equation}\label{lambda1}
\lambda_{n}({\bf x})=1+(\Lambda-1){\bf H}_{n}({\Omega}_s),
\end{equation}
where ${\bf H}_n$ is an approximation for the Heaviside function
\begin{equation}\label{lambda2}
{\bf H}_n({\Omega}_s)=\left\{
\begin{array}{rl}
&1~~~~~{\rm if}~~~~~{\bf x}\in {\Omega}_s,\\
&0~~~~~ {\rm otherwise},
\end{array} \right.
\end{equation} 
where we define ${\bf H}_n$ as
\begin{equation}
{\bf H}_n({\Omega}_s)=  \int \delta_n({\bf x}-{\bf x}_{\partial \Omega_s})d{\bf x}  ,
\end{equation}
in which 
\begin{equation}
\delta_n({\bf x})=\frac{ n}{\sqrt{2\pi}}\exp{(-\frac{n^2}{2}{\bf x}^2)},
\label{delta}
\end{equation}
which is one of the conventional ways in definition of the Heaviside function and Dirac delta function. Moreover, for the future use, we define a limiting case
\begin{equation}
\lambda_\infty=\lim_{n\rightarrow\infty}\lambda_n, \label{lambda_inf}
\end{equation}
which is the sharpest Heaviside function that we will use.\\
With regards to the above definitions the following points should be noticed:  
\begin{itemize}
\item[{\bf (i)}] The original Navier--Stokes equations (\ref{e0}) are valid in both $\Omega_f$ and $\Omega_s$; however, in $\Omega_s$ we have $\lambda\rightarrow \infty$,which means time is stopped. In particular, it means ${\bf u}(\Omega_s,t)={\bf u}^*(\Omega_s,t)/\lambda\rightarrow 0$ and  $\mu(\Omega_s,t)=\mu^*/\lambda\rightarrow 0$; that is, the $\Omega_s$ is occupied by an inviscid fluid with zero velocity.
\item[{\bf (ii)}] In the fluid region $\Omega_f$ where $\lambda=1$, we have ${\bf u}(\Omega_f,t)={\bf u}^*(\Omega_f,t)$, and the original Navier--Stokes equations (\ref{e0}) are in action. This is the region that the solution is sought in.
\item[{\bf (iii)}] In the interface region ${\rm supp}(\nabla\lambda)$ the forcing terms ${\bf F}^{\lambda}$ and ${\bf G}^{\lambda}$ are non-zero and the modified equations (\ref{e7})--(\ref{e8}) differ from the original Navier--Stokes equations. 
\end{itemize}
Naturally, in the interface region ${\rm supp}(\nabla \lambda)$ the behavior of the modified equations (\ref{e7})--(\ref{e8}) is under the influence of ${\bf F}^{\lambda}$ and ${\bf G}^{\lambda}$, which is discussed in the sequel.
\subsection{The physical and mathematical characteristics of the modified equations}\label{modif} 
The mathematical and physical properties of the modified equations (\ref{e7})--(\ref{e8}) are discussed in order to show that the dynamics of the inside and outside of the solid surface are separated. 
\subsubsection{The penalizing functions $\nabla \lambda/\lambda$ and $\nabla^2\lambda/\lambda$}\label{sec_delta}
The interface region $\Gamma_s$ affects the modified Navier--Stokes equations via the derivatives of $\lambda$, that is,  $\nabla \lambda/\lambda$ and $\nabla^2\lambda/\lambda$ (see the forcing functions (\ref{e8})). Therefore, the properties of the modified equations (\ref{e7})--(\ref{e8}) cannot be fully understood without insight on the properties of these penalizing functions. 

At first, according to the definition of $\lambda$, obviously
\begin{equation}
\lim_{\Gamma_s\rightarrow \partial\Omega_s} \frac{\nabla \lambda}{\vert \nabla \lambda \vert}=-\hat{\bf n},
\end{equation}
where $\vert \nabla \lambda \vert$ is the magnitude of $\nabla \lambda$, and $\hat{\bf n}$ is the unit normal vector of $\partial \Omega_s$. Therefore
\begin{equation}
\frac{\nabla \lambda}{\lambda}=-\frac{\vert \nabla \lambda \vert}{\lambda}\hat{\bf n}.
\end{equation} 
Since $\lambda$ is a positive number, one can conclude that $\nabla \lambda/\lambda$ is a vector, perpendicular to $\partial\Omega_s$ and in the direction of $-\hat{\bf n}$ (into $\Omega_s$). 

Now, to see the other properties of the penalizing terms, we investigate  $\lambda_n^{\prime}/\lambda_n$ and $\lambda_n^{\prime\prime}/\lambda_n$ on the $-\hat{\bf n}$ direction (denotes by $x$ coordinate), instead of investigating  $\nabla \lambda/\lambda$ and $\nabla^2\lambda/\lambda$ in the Cartesian coordinate.

The time dilation factor $\lambda_n$ is defined by Eqns. (\ref{lambda1})--(\ref{delta}). Now, assuming $(\Lambda-1)\approx \Lambda$ for $\Lambda\gg 1$, for a point ${x}_\Gamma$ on $\Gamma_s$ one can write:
\begin{eqnarray}
\lambda_n^\prime &=&\frac{d }{d x}(1+\Lambda{\bf H}_n)=\Lambda \delta_n=\Lambda \left( \frac{n}{\sqrt{2\pi}}\right)\exp{\left[-\frac{n^2}{2}(x-{ x}_\Gamma)^2\right]},\\
\lambda_n^{\prime\prime}&=&\frac{d}{d x}\lambda_n^\prime=\Lambda\frac{d\delta_n}{dx}=-\Lambda \left(\frac{n^3}{\sqrt{2\pi}}\right) (x-{ x}_\Gamma)\exp{\left[-\frac{n^2}{2}(x-{x}_\Gamma)^2\right]}.
\end{eqnarray}
Investigation of  these terms is in order: 
 
\begin{figure}
\centering
\includegraphics[width=1.\textwidth]{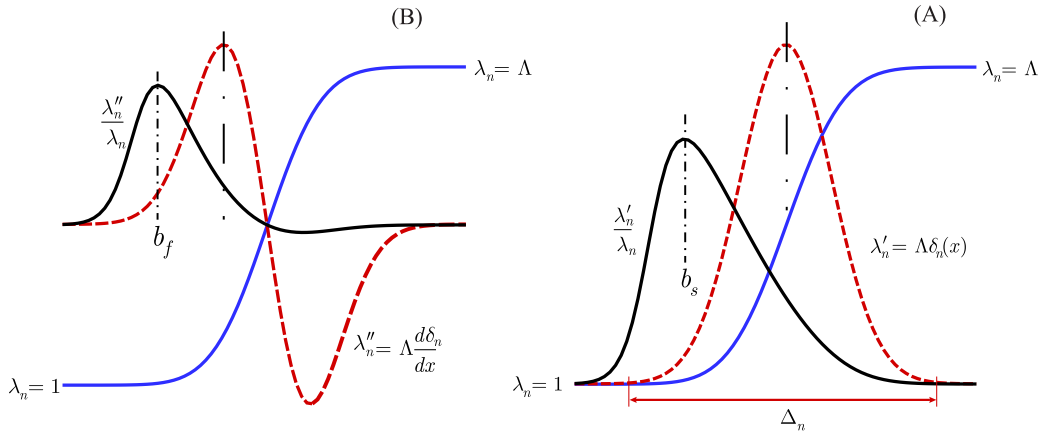}
\caption{The penalizing functions $\nabla \lambda/\lambda$ and $\nabla^2\lambda/\lambda$ in one space dimension. Left: $\lambda^{\prime\prime}/\lambda$. Right: $\lambda^{\prime}/\lambda$. In contrast to $\lambda^{\prime}$ and $\lambda^{\prime\prime}$ which are symmetric, $\lambda^{\prime\prime}/\lambda$ and $\lambda^{\prime}/\lambda$ are non-symmetric and are shifted to the fluid region. Although $\lambda^{\prime\prime}/\lambda\gg \lambda^{\prime}/\lambda$, the curves are scaled such that be comparable visually.} \label{fig21}
\end{figure}
\begin{itemize}
\item[{\bf (i)}]At first, if we define a characteristic width $\Delta_n=\frac{6}{n}$ for $\lambda_n^{\prime}=\Lambda\delta_n(x)$ (see panel (A) of Fig. \ref{fig21}), we can obtain an estimation for the order of magntude of $\lambda_n^{\prime}/\lambda_n$ and $\lambda_n^{\prime\prime}/\lambda_n$. That is \footnote{The definition $\Delta_n=\frac{6}{n}$ has its root in the conventional definitions $6\sigma$ width for a normal distribution, where $\sigma$ is the standard deviation (we know that $\lambda_n(\pm 3\sigma)\approx 0.003\Lambda$).}
\begin{eqnarray}
{\mathcal O}\left(\frac{\lambda_n^{\prime}}{\lambda_n}\right) &=&{\mathcal O}\left(\frac{\frac{d }{d x}\lambda_n}{\lambda_n}\right)=
\frac{ \frac{\mathcal O \left( \frac{\Lambda}{\sqrt{2\pi}}  \right)}{\mathcal O \left( \Delta_n \right)}  }{ \mathcal O \left( \frac{\Lambda}{\sqrt{2\pi}}  \right) }={\mathcal O}\left( \frac{1}{\Delta_n}\right)=
{\mathcal O}\left(n\right),   \\  \nonumber \\ 
{\mathcal O}\left(\frac{\lambda_n^{\prime\prime}}{\lambda_n}\right)&=&{\mathcal O}\left(\frac{\frac{d^2}{d x^2} \lambda_n}{\lambda_n}\right)=
\frac{ \frac{\mathcal O \left( \frac{\Lambda}{\sqrt{2\pi}}  \right)}{\mathcal O \left( {\Delta_n}^2 \right)}  }{ \mathcal O \left( \frac{\Lambda}{\sqrt{2\pi}}  \right) }={\mathcal O}\left( \frac{1}{\Delta_n^2}\right)={\mathcal O}\left(n^2\right).
\end{eqnarray}
Both are large numbers and both go to infinity as $n$ goes to infinity (this is the reason that we recognize them as the penalizing functions); although $\lambda_n^{\prime\prime}/\lambda_n > \lambda_n^{\prime}/\lambda_n$.  Moreover, note that the sharpness of the interface (i.e., $\Delta_n$) is controled via $n$.
\item[{\bf (ii)}] In contrast to $\lambda^{\prime}$ and $\lambda^{\prime\prime}$ which are symmetric, $\lambda_n^{\prime}/\lambda_n$ and $\lambda_n^{\prime\prime}/\lambda_n$ have not symmetry axes. This issue is emphasised in Fig. \ref{fig21}. In panel (A), $\lambda_n$, $\lambda^\prime_n$, and $\lambda^\prime_n/\lambda_n$ are illustrated. As one can see, $\lambda^\prime_n/\lambda_n$ is skewed and its maximum is shifted to the left (to the fluid region). The same issue can be seen in panel (B) in comparison of $\lambda_n^{\prime\prime}$ and $\lambda_n^{\prime\prime}/\lambda_n$. Again  $\lambda_n^{\prime\prime}/\lambda_n$ is skewed and its maximum value is shifted to left (i.e. to the fluid region). The places that  $(\lambda^\prime_n/\lambda_n)_{\rm max}$ and $(\lambda_n^{\prime\prime}/\lambda_n)_{\rm max}$ are occurred will be called $b_f$ and $b_s$.   
\item[{\bf (iii)}] $(\lambda_n^{\prime}/\lambda_n)_{\rm max}$ and $(\lambda_n^{\prime\prime}/\lambda_n)_{\rm max}$ occur in different places for finite $n$. To stress the issue $\lambda_n^{\prime}/\lambda_n$ and $\lambda_n^{\prime\prime}/\lambda_n$ are illustrated in Fig. \ref{fig22}. It is noticable that $b_f$ is closer to the fluid region, and $b_s$ is deeper in the solid region. As it is shown in Fig.  \ref{fig22} the distance between $b_f$ and $b_s$ is called $\Delta_n^\prime$. For this particular form of $\delta_n$ that we used (see Eq. (\ref{delta})), one can show that $\Delta_n^\prime\approx0.125\Delta_n$. However, note that $\Delta_n$ and $\Delta^\prime_n$ both go to zero as $n$ goes to infinity. 
\end{itemize}
Now we are ready to investigate the mathematical and physical characteristics of the modified equaitions.
\begin{figure}
\centering
\includegraphics[width=0.7\textwidth]{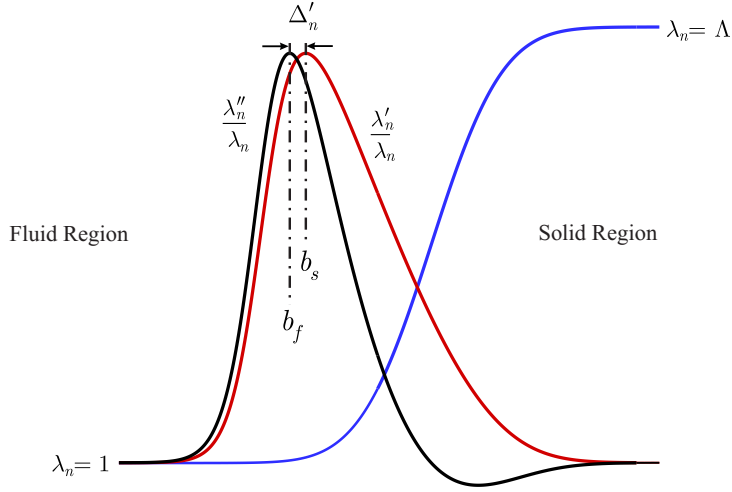}
\caption{The maximums of $\lambda^{\prime}/\lambda$ and $\lambda^{\prime\prime}/\lambda$ occur in different places, and $b_f$ is closer to the fluid region. $\Delta_n^\prime\approx0.125\Delta_n$, therefore, by approaching $n$ to infinity, $\Delta_n^\prime\rightarrow 0$, and $b_f$ and $b_s$ approach to each other.} \label{fig22}
\end{figure}
\subsubsection{The modified continuity equation} \label{continuity}
Consider the modified continuity equation
\[
\nabla\cdot{\bf u}-{\bf G}^{\lambda}=\nabla\cdot{\bf u}+\frac{1}{\lambda}{\bf u}^T\cdot \nabla\lambda=0.
\]

This particular form of the mass conservation law (or the volume conservation, since $\varrho$ is constant) is a result of curvature of the space--time manifold. In fact, in the presence of time dilation, the net spatial rate of change of velocity of the fluid particles can be written as
\begin{equation}
\frac{\partial }{\partial x}(\lambda u)=\lambda \frac{\partial u }{\partial x}+u \frac{\partial \lambda}{\partial x},
\end{equation}
which means the net rate of change is a summation the rate of change of velocity in a constant $\lambda$ and the rate of change due to change of $\lambda$, that is, the curvature in space--time manifold. Now, summation in all directions and equating by zero results in the modified continuity equation. 

However, note that the above equations were derived for a general $\lambda$. But, when $\lambda\rightarrow \lambda_\infty$ (see Eq. (\ref{lambda_inf})), we have ${\rm supp}(\nabla \lambda)\rightarrow \partial \Omega_s$, where ${\bf u}=0$ (as it will be seen in the next section $\S$ \ref{moment1}) which means 
\begin{equation}
{\rm{\bf G}^\lambda}=0, \quad \quad {\rm everywhere~in}~~ \bar{\Omega}
\end{equation}
and the classical continuity equation is retrieved.
  
In fact, when $\lambda\rightarrow \lambda_\infty$, there is no any flow across the solid boundary; there are two separate regions (i.e., the fluid region and the solid region), each one has its constant $\lambda$, and therefore, $\nabla\cdot {\bf u}=0$ in both regions separately.   
\subsubsection{The modified momentum equation}\label{moment1}
Assuming that ${\bf u}({\bf x},t)$ is solenoidal everywhere in ${\Omega}$ at each time instant, the dynamics of the flow governs by the momentum equation. Therefore, in order to separate the dynamics of the inside and outside of the solid body, it should be no momentum exchange between two sides of the solid boundary $\partial\Omega_s$.  In the following we explain how the penalization terms Eq. (\ref{e8}) do this indeed.

In order to facilitate the analysis, we decompose the forcing function ${\bf F}^\lambda$ (see Eq. (\ref{e8})) to
\begin{equation}
{\bf F}^{\lambda}={\bf F}_{\rm Diff.}^{\lambda}+{\bf F}_{\rm Adv.}^{\lambda}+{\bf F}_{\rm Press.}^{\lambda}
\label{F_tot}
\end{equation}
where
\begin{eqnarray}
{\bf F}_{\rm Diff.}^{\lambda}&=&\frac{\nu}{\lambda} ({\bf u}\nabla^2\lambda+2\nabla {\bf u}\cdot \nabla \lambda), \label{Diff}\\
{\bf F}_{\rm Adv.}^{\lambda}&=&-\frac{1}{\lambda} {\bf u}({\bf u}^T\cdot \nabla)\lambda,\label{Conv} \\
{\bf F}_{\rm Press.}^{\lambda}&=& -\frac{2}{\lambda}\cdot \frac{p}{\varrho}\nabla \lambda.\label{Press}
\end{eqnarray}
\noindent
Particularly, it is aimed to show: 
\begin{enumerate}
\item There is no diffusion between two sides of the solid boundary.
\item There is no advection between two sides of the solid boundary.
\item The pressure acts separately in each region, that is, the pressure of each region does not affect the other region.
\end{enumerate}
Below these issues are discussed.\\

\begin{figure}
\centering
\includegraphics[width=1.1\textwidth]{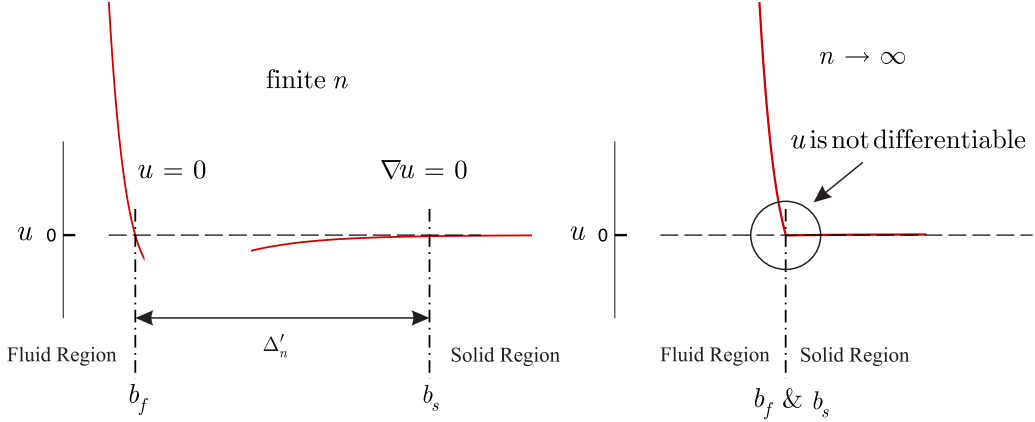}
\caption{Simultaneous imposition of the no-slip and no-diffusion conditions make the velocity ${\bf u} \in C^0$. Left: the no-slip condition enforces ${\bf u}|_{b_f}=0$ while the no-diffusion condition enforces $\nabla{\bf u}|_{b_s}=0$. Right: when $n\rightarrow\infty$, then $\Delta^\prime_n\rightarrow 0$, therefore, $b_f$ and $b_s$ approach to each other, and ${\bf u}$ approaches to a continuous function that is not differentiable at the boundary.} \label{fig5}
\end{figure}
\noindent
{\bf (1) No diffusion between two sides of the solid boundary}

The diffusion forcing function 
\[
{\bf F}_{\rm Diff.}^{\lambda}=\nu{\bf u}\frac{\nabla^2\lambda}{\lambda} +2\nu\nabla {\bf u}\cdot \frac{\nabla \lambda}{\lambda},
\]
 is obtained directly from expansion of the viscous (diffusion) terms of the Navier--Stokes equations. In fact, we expanded $\frac{1}{\lambda^2}(\lambda\nu)\nabla^2(\lambda {\bf u})$, and used the vector identities (\ref{identity}) to obtain  ${\bf F}^\lambda_{\rm Diff.}$.

Now, note that ${\bf F}^\lambda_{\rm Diff.}$ has two terms, in the first one ${\bf u}$ is multiplied by the penalizing term $\nabla^2 \lambda/\lambda$, and  in the second one $\nabla {\bf u}$ is multiplied by $\nabla \lambda/\lambda$. Therefore, we have both ${\bf u}\rightarrow 0$ and $\nabla{\bf u}\rightarrow 0$ at the boundary, which means: 
\begin{itemize}
\item[{\bf (i)}] Obviously, ${\bf u}=0$ means no diffusion due to transfer of mass; the term that we usually refer to as the no-slip condition.
\item[{\bf (ii)}] Recalling the classical equation for the diffusion of momentum, that is $\nabla\cdot(\nu\nabla{\bf u})$ , one can see that, $\nabla{\bf u}=0$ means no diffusion of momentum due to the velocity gradient. 
\end{itemize}
Therefore, all kinds of diffusion between two sides of the solid boundary are discarded by letting ${\bf u}\rightarrow 0$ and $\nabla{\bf u}\rightarrow 0$ at the boundary.\\

There is still another interesting fact which will be revealed when we notice to the positions that the no-slip condition ${\bf u}=0$ and the no-diffusion condition  $\nabla{\bf u}= 0$ are occurred: 
\begin{quote}
${\bf u}$ is multiplied by $\nabla^2 \lambda/\lambda$ while $\nabla{\bf u}$ is multiplied by $\nabla \lambda/\lambda$. Therefore, recalling our discussion in section \ref{sec_delta}, one can say that ${\bf u}=0$ is occurred at $b_f$ while $\nabla{\bf u}=0$ is occurred at $b_s$. Therefore, the velocity ${\bf u}({\bf x})$ is so that ${\bf u}|_{b_f}=0$ and $\nabla{\bf u}|_{b_s}=0$. Now, when $\lambda\rightarrow\lambda_\infty$,we have $\Delta_n\rightarrow 0$ and $\Delta^\prime_n\rightarrow 0$. Therefore,  $b_f$ and $b_s$ approach to each other, making ${\bf u}({\bf x})\in{\mathcal C}_0 $. It is particularly in agreement with our anticipation of velocity in the presence of a solid body. The issue is illustrated intuitively in Fig. \ref{fig5}.
\end{quote}

\noindent
{\bf (2) No advection between two sides of the solid boundary}

The advection penalization term
\[
{\bf F}_{\rm Adv.}^{\lambda}=-{\bf u}({\bf u}^T\cdot \frac{\nabla\lambda}{\lambda}),
\]
is obtained from expansion of the advection term of the Navier--Stokes equations. In fact, we expanded $\frac{1}{\lambda^2}[(\lambda{\bf u})^T\cdot\nabla](\lambda{\bf u})$, and used the vector identities (\ref{identity}) to obtain ${\bf F}^\lambda_{\rm Adv.}$. 

As one can see, ${\bf u}{\bf u}^T$ is multiplied by $\frac{\nabla \lambda}{\lambda}$. Therefore, this term enforces ${\bf u}{\bf u}^T\rightarrow 0$, at $b_s$, that is, the no-momentum flux condition. It should be emphasized that although the last no-diffusion condition imposes $\nabla {\bf u}=0$ at $b_s$, it does not guarantee ${\bf u}=0$ at $b_s$. In fact,  the no-momentum flux is another condition which is somehow independent of the no-diffusion condition.\\

\noindent
{\bf (3) No pressure coupling between inside and outside of the solid body}

In analysis of the momentum equation, we assume that the pressure is a known quantity (obtained somehow from an elliptic equation, to impose the continuity). Therefore, the pressure is not going to be determined directly in the momentum equation. As a consequence, the decoupling of the pressure cannot be fully justified here by merely analysis of the momentum equation without considering the elliptic pressure equation (which will be discussed in the next section). 

However, with regard to the pressure penalization term ${\bf F^\lambda_{\rm Press.}}$ the following issues are noticeable:
\begin{figure}[t]
\centering
\includegraphics[width=0.55\textwidth]{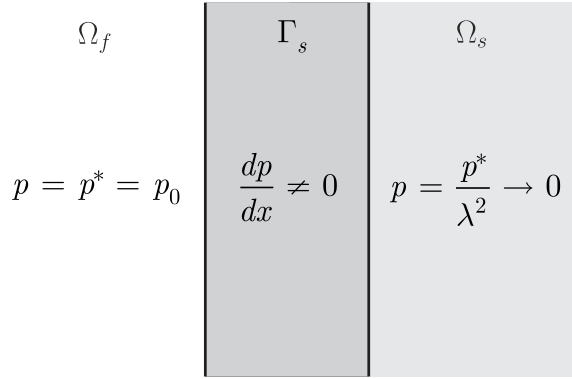}
\caption{The role of ${\bf F^\lambda_{\rm Press.}}$. We assumed that there is a constant pressure $p_0$ in the fluid region, while $p=\frac{p_0}{\lambda^2}\rightarrow 0$ in the solid region. The non-zero pressure gradient in the interface region $\Gamma_s$ should be compensated such that $\frac{dp}{dx}$ do not induce a net driving force to the fluid particles, this is the way that ${\bf F^\lambda_{\rm Press.}}$ affects the momentum equation.} \label{fig6}
\end{figure}
\begin{itemize}
\item[{\bf (i)}] The pressure penalization term
\[
{\bf F}_{\rm Press.}^{\lambda}= -2\frac{p}{\varrho}\frac{\nabla \lambda}{\lambda},
\]
is obtained from expansion of the pressure gradient term of the Navier--Stokes equations. In fact, we expanded $-\frac{1}{\lambda^2}\frac{1}{\varrho^*}\nabla(\lambda^2p)$ and used the vector identities (\ref{identity}) to obtain ${\bf F^\lambda_{\rm Press.}}$. 

It is well-known that in an incompressible fluid flow the pressure driving force produces by the pressure gradient not by the absolute value of the pressure (the reason that the pressure gradient appears in the momentum equation). With this regard, the forcing term $-2\frac{p}{\varrho}\frac{\nabla \lambda}{\lambda}$ in which $p$ is appeared explicitly is somehow unusual, and must be justified.
\item[{\bf (ii)}] To elucidate the issue, we consider the hydrostatic pressure in the vicinity of a solid boundary (see Fig. \ref{fig6}). To simplify the situation, we assume that all the velocities are zero, and we neglect the external forces. Therefore, we are faced with a constant hydrostatic pressure, say $p_0$, in the fluid. Now, notice to the pressure in the vicinity of the solid boundary in Fig. \ref{fig6}. We have a pressure jump at the boundary, because the pressure changes from $p=p^*=p_0$ in the fluid to $p=\frac{p^*}{\lambda^2}\rightarrow 0$ in the solid body. Therefore, the pressure encounters a non-zero gradient in $\Gamma_s$. However, this pressure gradient should not impose an acceleration to the fluid particles indeed, because it is a result of curvature in the space--time manifold. Therefore, this pressure gradient must be balanced by another term, which is ${\bf F}^\lambda_{\rm Press.}$. 

In order to show that ${\bf F}^\lambda_{\rm Press.}$ balance the pressure gradient in crossing the boundary, we write the modified momentum equation for this flow. All the velocities are zero, and there is no external forces, therefore, from the modified momentum equation
\begin{equation}
-\frac{1}{\varrho}\nabla p-2\frac{p}{\varrho}\frac{\nabla \lambda}{\lambda}=0 \quad \Longrightarrow \quad -\frac{1}{\varrho}\nabla p=2\frac{p}{\varrho}\frac{\nabla \lambda}{\lambda},
\end{equation}
which means the pressure gradient is balanced by ${\bf F^\lambda_{\rm Press.}}$ everywhere in ${\Omega}$, including at the solid boundary. But ${\bf F^\lambda_{\rm Press.}}=0$ everywhere except at the solid boundary.   
\end{itemize} 
\subsubsection{On the conservation of ${\bf F^\lambda}$}
Before ending analysis of the modified momentum equation, we shall investigate solenoidal property of ${\bf F^\lambda}$. To this end, we introduce $\tilde{\bf F}^\lambda$ as
\begin{equation}
\tilde{\bf F}^\lambda={\bf F}_{\rm Diff.}^{\lambda}+{\bf F}_{\rm Adv.}^{\lambda},
\end{equation}
and according to Eq. (\ref{F_tot}), we have
\begin{equation}
{\bf F^\lambda}=\tilde{\bf F}^\lambda+{\bf F^\lambda_{\rm Press.}}.
\end{equation}
Now it is not so difficult to verify that 
\begin{equation}
\nabla\cdot \tilde{\bf F}^\lambda=0,
\end{equation}
that is, $\tilde{\bf F}^\lambda$ is conservative. We will use this property in obtaining the pressure elliptic equation in the next section.
\subsubsection{Elliptic nature of the flow and the pressure equation}
Isolating a part of the solution domain in an elliptic equation is one of the most difficulties of any theory of interaction of a solid body with an incompressible flow. The theory should allow pressure jump at the boundary, and the pressure of inside of the solid body should not affect the pressure of the outside, and vice versa. In the present method, the pressure jumps to zero at the solid boundary, and accordingly, in the elliptic pressure equation, presence of penalizing terms in the left hand side allow this jump and vanish the pressure gradient in the solid body. 

Similar to the classical Navier--Stokes equations, here we derive the pressure equation by taking divergence of the modified momentum equations, that is
\begin{equation}
\nabla\cdot(\partial_t {\bf u})+\nabla\cdot\left[({\bf u}^T\cdot \nabla){\bf u}\right]=-\nabla\cdot\left[\frac{1}{\lambda^2\varrho^*}\nabla p^*\right]+\nu^*\nabla\cdot\left[\frac{\nabla^2{\bf u}}{\lambda}\right]+\nabla\cdot(\tilde{{\bf F}}^{\lambda}).
\end{equation}   
Note that ${\bf F^\lambda_{\rm Press.}}$ is included in the pressure gradient term, therefore, the forcing term in the right hand side is $\tilde{\bf F}^\lambda$ not ${\bf F}^\lambda$; moreover, we substituted $\nu=\nu^*/\lambda$. Now, since ${\bf u}$ and $\tilde{{\bf F}}^{\lambda}$ are divergence free, by simplifying the equation and substitution of $\varrho=\varrho^*$ we have
\begin{equation}
\nabla\cdot\left(\frac{1}{\lambda^2}\nabla p^*\right)=-\varrho \nabla\cdot[({\bf u}^T\cdot\nabla){\bf u}]+\mu \frac{\nabla \lambda}{\lambda}\nabla^2{\bf u}.
\label{PPE}
\end{equation}
Now, notice to the left hand side of the equation. This is a diffusion equation for $p^*$ with a variable diffusion coefficient $\frac{1}{\lambda^2}$. In the fluid region $\frac{1}{\lambda^2}=1$ and the classical Poisson's pressure equation is retrieved, while in the solid region $\frac{1}{\lambda^2}\rightarrow 0$, and $p^*={\rm Cte.}$, irrespective of the right hand side of the equation. Therefore, one can say
\begin{equation}
p=\frac{p^*}{\lambda^2}\rightarrow 0 \quad \quad x\in \Omega_s.
\end{equation}
This property of the pressure will be seen in practice in our exact solutions (in section \ref{solutions}).

Although the present form of Eq. (\ref{PPE}) is physically meaningful, it is not in the familiar form of the elliptic equations. In order to convert it to a more conventional form, we expand the left hand side. In this manner
\begin{equation}
\nabla^2 p+2\frac{\nabla \lambda}{\lambda}\nabla p+2 \nabla\cdot \left( \frac{\nabla \lambda}{\lambda}\right) p=-\varrho \nabla\cdot[({\bf u}^T\cdot\nabla){\bf u}]+\mu \frac{\nabla \lambda}{\lambda}\nabla^2{\bf u}.
\label{PPE2}
\end{equation}
This is the elliptic pressure equation, and plays the role of the Poisson's pressure equation in the classical Navier--Stokes equations.
\section{Application of the modified equations: (1) Exact solutions} \label{solutions}
As it was discussed above, all the mechanisms of momentum exchange between the solid body and the fluid flow are discarded by the penalization terms at the solid boundary. Therefore, exact imposition of a solid body on the solution of the incompressible Navier--Stokes equations is anticipated.   
\subsection{The Stokes first problem}
Since both the no-slip and no-diffusion conditions are satisfied by the penalization terms, a diffusive boundary can be implemented exactly. 
\subsubsection{An overview on the classical solution}
In a uniform two-dimensional velocity field $U$ on the half plan $(x,y\geq 0)$ the no-slip condition is implemented at $y=0$ from $t=0^+$. By simplification of the classical Navier--Stokes equations, we have:
\begin{equation}
\frac{\partial u}{\partial t}=\nu \frac{\partial^2 u}{\partial y^2}. \label{Stokes0}
\end{equation}
Now, by definition of the similarity variables
\begin{equation}
\eta=\frac{y}{2\sqrt{\nu t}}; \quad {\rm and} \quad f_{\rm Stokes} (\eta)=\frac{u}{U},
\end{equation}
and substitution in Eq. (\ref{Stokes0}), one obtains
\begin{equation}
\left\{
\begin{array}{rl}
&f_{\rm Stokes}^{\prime\prime}+2\eta f_{\rm Stokes}^\prime=0, \\
&f_{\rm Stokes}(\eta =0)=0,\\
&f_{\rm Stokes}(\eta \rightarrow \infty)=1,
\end{array} \right. \label{Stokes3}
\end{equation}
with the exact solution of 
\begin{equation}
f_{\rm Stokes}(\eta)={\rm erf}(\eta)=\frac{2}{\sqrt{\pi}}\int_{0}^{\eta}e^{-x^2}dx, \label{Exact1}
\end{equation}
on the interval $\eta \in [0,+\infty)$. The primes show derivatives with respect to $\eta$.

\subsubsection{Set up of the new solution}
\begin{figure}[t]
\centering
\includegraphics[width=0.8 \textwidth]{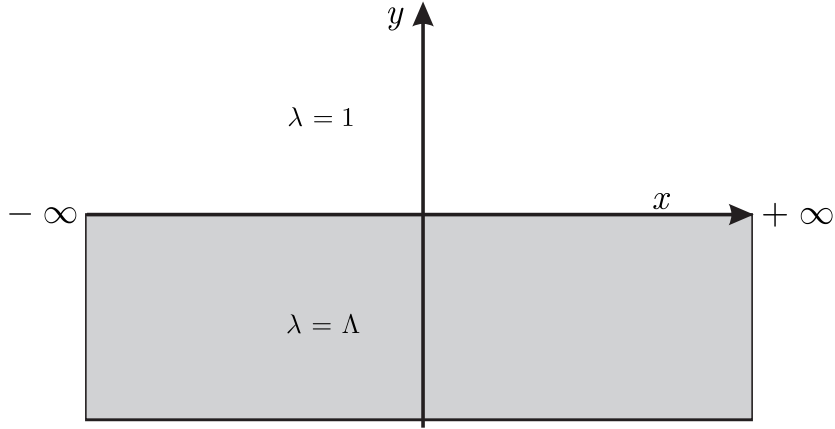}
\caption{Set up of the Stokes first problem using the time dilation. From the time instant $t=0^+$, time dilation $\lambda=\Lambda$ is implemented in the lower half plane $y\leq 0$, where a uniform flow $U$ is existed.} \label{fig7}
\end{figure}

According to Fig. \ref{fig7}, in a uniform two-dimensional velocity field $U$, a time dilation $\lambda=1+\Lambda {\rm {\bf H}}(y)$ is imposed from $t=0^+$, where ${\rm {\bf H}}(\cdot)$ is the Heaviside function and $\Lambda\rightarrow \infty$ is a large positive number. 

First, we simplify the modified Navier--Stokes equations for this flow. 
Beginning from the modified continuity equation (\ref{}) (recalling that ${\rm {\bf G}}^\lambda=0$), since $\frac{\partial}{\partial x}=0$, we have
\begin{equation}
\frac{\partial v}{\partial y}=0. \Longrightarrow v(x,y)=0.
\end{equation}  
Now, using $v(x,y)=0$, the modified $y$--momentum equation (ignoring the external forces) reduces to
\begin{equation}
\frac{1}{\lambda^2}\frac{d}{dy}(\lambda^2 p)=0 \Longrightarrow \lambda^2 p={\rm Cte.}
\end{equation} 
Assuming a constant pressure in the fluid region, say $p_0$, one can write
\begin{equation}
p(\Omega_s)=p(y<0)=\frac{p_0}{\lambda^2}\rightarrow 0,
\end{equation}
which means the pressure of the solid region is zero.\\
Now, from the modified $x$--momentum equation
\begin{equation}
\frac{\partial u}{\partial t}=\nu \frac{\partial^2 u}{\partial y^2}+\underbrace{\nu\left[\left(\frac{1}{\lambda}\frac{\partial^2 \lambda}{\partial y^2}\right)u+2\left(\frac{1}{\lambda}\frac{\partial \lambda}{\partial y} \right) \frac{\partial u}{\partial y} \right]}_{F^\lambda _x}. \label{Stokes1}
\end{equation}
This is the governing equation both for the solid and fluid regions, and as it will be seen it can be solved by the similarity solution.
\subsubsection{Similarity equation}
We chose the local similarity variable as
\begin{equation}
\eta^*=\frac{y}{2\sqrt{\nu^* t^*}}.
\end{equation}
But since $t^*=t/\lambda$ and $\nu^*=\lambda \nu$, then
\begin{equation}
\eta^*=\eta=\frac{y}{2\sqrt{\nu t}}.
\end{equation}
And for the similarity function 
\begin{equation}
 f^*(\eta^*)=\frac{u^*}{U^*}=\frac{u}{U}=f(\eta).
\end{equation}
By substitution in Eq. (\ref{Stokes1}), we faced with the problem
\begin{equation}
\left\{
\begin{array}{rl}
&f^{\prime\prime}+2\eta f^\prime+2\frac{\lambda^\prime}{\lambda}f^\prime+\frac{\lambda^{\prime\prime}}{\lambda}f=0,\\ \\
&f(\eta \rightarrow - \infty)=0,\\
&f(\eta \rightarrow + \infty)=1,
\end{array} \right. \label{Stokes2}
\end{equation}
which should be solved to find the solution in $\eta \in (-\infty,+\infty)$. In this equation primes stand for the derivatives with respect to $\eta$. With regard to this problem, the following points should be noted:\\
\begin{itemize}
\item[{(1)}] There is a forcing term ${F}_x^\lambda=-2\frac{\lambda^\prime}{\lambda}f^\prime-\frac{\lambda^{\prime\prime}}{\lambda}f$  in Eq. (\ref{Stokes2}) with the support of ${\rm supp}(\lambda^\prime)$ (i.e., $\eta=0$) which imposes both the no-slip condition  $f=0$ and no-diffusion condition $f^\prime=0$. 
\item[{(2)}] The classical Stokes first problem (\ref{Stokes3}) is defined on $\eta \in [0,+\infty)$, and the solid boundary is implemented by imposing the no-slip condition at $\eta=0$; while in our problem (\ref{Stokes2}) the solution domain is $\eta \in (-\infty,+\infty)$, and the solid boundary is implemented in $\eta = 0$ via the forcing term ${F}_x^\lambda$ (see Fig. \ref{fig8}).
\item[{(3)}] With the best knowledge of the author, this is the first time that a single similarity equation is found for a fluid--solid system.
\end{itemize}
In the sequel, at first we find the exact solution of Eq. (\ref{Stokes2}) and then we will solve it numerically.
\subsubsection{Exact solution}
A closed-form solution can be obtained for problem (\ref{Stokes2}). To this end, at first, we extend the Stokes solution to the negative $\eta$. Obviously, this extension is not unique. However, for the present problem, the simplest way is extending the range of $\eta$ 
\begin{equation}
f^{\rm Ext}_{\rm Stokes}(\eta)=\frac{2}{\sqrt{\pi}}\int_{0}^{\eta}e^{-x^2}dx, \quad \quad \eta\in (-\infty,+\infty).
\end{equation}
Now, for the exact solution $f_e$ we suggest:
\begin{equation}
f_{e}(\eta)=G(\eta)\cdot f^{\rm Ext}_{\rm Stokes}(\eta)=G(\eta)\frac{2}{\sqrt{\pi}}\int_{0}^{\eta}e^{-x^2}dx,
\label{f_1}
\end{equation}
where $G(\eta)$ is a function that is going to be found such that $f_e$ satisfy Eq. (\ref{Stokes2}). From Eq. (\ref{f_1}) and using the Leibnitz rule, one can obtain:
\begin{equation}
f_e^\prime = \underbrace{G^\prime \frac{2}{\sqrt{\pi}}\int_{0}^{\eta}e^{-x^2}dx}_{\mathcal {I}}+G\frac{2}{\sqrt{\pi}}e^{-\eta^2}.
\end{equation}
$G^\prime$ is required to be non-zero only in the vicinity of $\eta=0$. But in this place, $\int_{0}^{\eta=0}e^{-x^2}dx=0$, therefore, term $\mathcal {I}$ is vanished, and
\begin{equation}
f_e^\prime =\frac{2}{\sqrt{\pi}} G(\eta) e^{-\eta^2}.
\label{f_2}
\end{equation}
And by differentiation once
\begin{equation}
f_e^{\prime\prime} =\frac{2}{\sqrt{\pi}} \left( G^{\prime}-2 \eta G \right) e^{-\eta^2}.
\label{f_3}
\end{equation}
Now, by substitution of Eqns.  (\ref{f_1}), (\ref{f_2}) and (\ref{f_3}) in Eq. (\ref{Stokes2}), one obtains
\begin{equation}
\left( G^\prime+2\frac{\lambda^\prime}{\lambda}G \right)e^{-\eta^2}+\underbrace{\frac{\lambda^{\prime\prime}}{\lambda}G\int_{0}^{\eta}e^{-x^2}dx}_{\mathcal{II}}=0.
\end{equation}
As we kow, $\frac{\lambda^{\prime\prime}}{\lambda}$ is non-zero only in the vicinity of $\eta=0$, where $\int_{0}^{\eta=0}e^{-x^2}dx=0$. Therefore, term $\mathcal{II}$ vanishes, and we have a first--order ODE
\begin{equation}
\left\{
\begin{array}{rl}
{\rm O.D.E.:} &G^\prime+2\frac{\lambda^\prime}{\lambda}G=0,\\ \\
{\rm B.C.:~~} &G(\eta >0)=1.
\end{array} \right. \label{f_4}
\end{equation}  
The B.C. is obtained by the fact that we want $f_e$ meet $f_{\rm Stokes}$ for $\eta>0$.\\
Integrating the above equation yields:
\begin{equation}
G(\eta)=\frac{1}{\lambda^2(\eta)}.
\end{equation}
Therefore, Eq. (\ref{Stokes2}) has a closed-form solution 
\begin{equation}
f_e(\eta)=\frac{2}{\sqrt{\pi}}\frac{1}{\lambda^2}\int_{0}^{\eta}e^{-x^2}dx,   \quad \quad \eta\in (-\infty,+\infty).
\label{Exact1_new}
\end{equation}
As one can see
\begin{equation}
f_e(\eta)=\left\{
\begin{array}{rl}
&f_{\rm Stokes}~~~~~{\rm for}~~~~~\eta> 0,\\
&0~~~~~~~~~~~{\rm for}~~~~~\eta< 0.
\end{array} \right.
\end{equation}
In order to have an estimation for the error in $\eta\gtrapprox 0$, we define the error as
\begin{equation}
{\rm Err}=\int_0^{+\infty} \vert f_e-f_{\rm Stokes} \vert d\eta=\int_0^{\Delta} \vert f_e-f_{\rm Stokes} \vert d\eta,
\end{equation}
where $\Delta$ is the characteristic width of ${\bf H}(\eta)$. Now, by substitution
\begin{equation}
{\rm Err}=\int_0^{\Delta} \vert  \underbrace{(1-\frac{1}{\lambda^2})}_{I}\left( \underbrace{\frac{2}{\sqrt{\pi}}\int_{0}^{\eta}e^{-x^2}dx}_{II}   \right)  \vert  d\eta
\end{equation}
where $I$ can be approximated as bellow:
\begin{equation}
I=(\frac{\lambda-1}{\lambda})(\underbrace{\frac{\lambda+1}{\lambda}}_{{\mathcal O}(1)})\approx (\frac{\lambda-1}{\lambda})\approx 1-\frac{\eta}{\Delta}.
\end{equation}
And for small $\Delta$, $II$ can be approximated as a linear function
\begin{equation}
II=\frac{2}{\sqrt{\pi}}\int_{0}^{\eta\leq \Delta}e^{-x^2}dx\approx a\eta,
\end{equation}
where $a$ is a constant. By substitution
\begin{equation}
{\rm Err}\leq \int_0^{\Delta}a\eta(1-\frac{\eta}{\Delta})d\eta=\frac{a}{6}\Delta^2,
\end{equation}
which means
\begin{equation}
{\rm Err}={\mathcal O}(\Delta^2).
\end{equation}
In addition to the exact solution, we shall solve Eq. (\ref{Stokes2}) numerically.  
\subsubsection{Numerical solution}
The Heaviside function ${\mathbf H}(\eta)$ is constructed by a scaled error function ${\rm erf}(\mathit{S}\eta)$, where $\mathit{S}$ is a scale factor which controls the sharpness of the Heaviside function. The derivatives $\lambda^\prime$ and $\lambda^{\prime\prime}$ are approximated by second order central finite--differencing of the Heaviside function. By $\lambda^\prime$ and $\lambda^{\prime\prime}$ in hand, equation (\ref{Stokes2}) is discretized on $\eta \in [-5,5]$, using the second--order finite difference method.  The discretized equation results in a tri-diagonal system of equations that is solved by the classical Thomas algorithm.  
\begin{figure}[t]
\centering
\includegraphics[width=1 \textwidth]{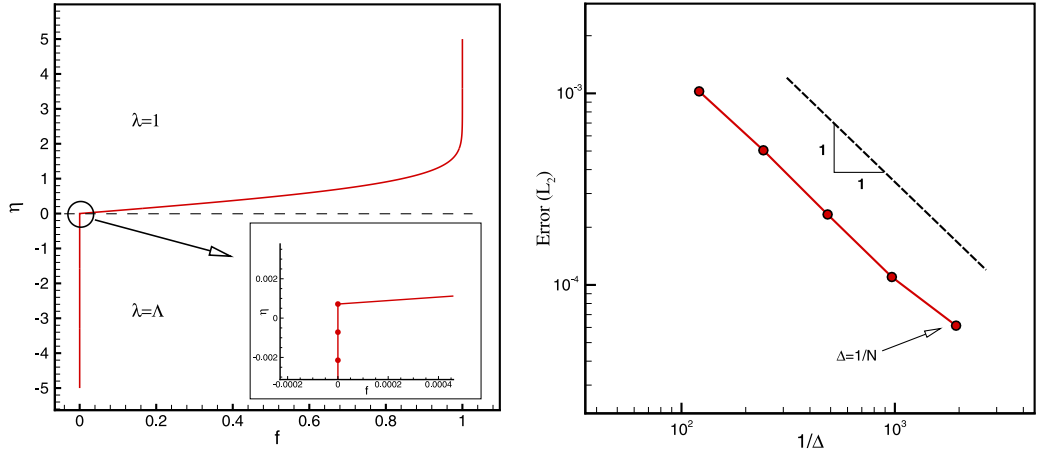}
\caption{The results of a second--order finite difference solution of Eq. (\ref{Stokes2}) on a $N=2048$ point uniform grid with $\Lambda=10^{30}$. Left: $f(\eta)$ on $-5 \leq \eta \leq 5$. On the solid domain $\eta\leq 0$ we have $f(\eta)=0$ while in the fluid domain $\eta\geq 0$, $f(\eta)$ follows the Stokes solution. Right: the rate of decaying of $L_2$ norm of the solution (in comparison with the exact solution) is plotted versus $\Delta$. The rate of decaying is obtained ${\mathcal O}(1)$ as it is anticipated.} \label{fig8}
\end{figure}

$\Lambda=10^{30}$ is chosen, and the problem is solved on a $N=2048$ point uniform grid with different $\Delta$. The results are shown in Fig. \ref{fig8}. In the left panel, $f(\eta)$  is shown on $\eta\in [-5,5]$ for $\Delta_{\rm min}\sim 1/N$. As one can see, $f(\eta)$ behaves like the error function for $\eta\geq 0$, while for $\eta \leq 0$, $f(\eta)$ is zero. Additionally, in this panel, a closer view of $f(\eta)$ in $\eta\approx 0$ is shown.  Obviously, the penalizing terms have managed to make $f(\eta)$ zero at the first grid point in the fluid region; and $\partial f/ {\partial} \eta=0$ at the first grid point in the solid region. The $L^2$ convergence rate of the solution versus $\Delta$ is in the right panel. The $\Delta$ is reduced until $\Delta_{\rm min}\sim 1/N$ (i.e., the Heaviside function width is one grid distance). As one can see, the convergence rate is first order as it is anticipated.  
\subsection{Plane stagnation point flow}
Although the method can easily be applied to the Falkner--Skan flow, as a wider class of problems, however, the stagnation point flow is chosen here,  because it is obtained directly from the Navier--Stokes equations not from a simplified version of them (i.e., the boundary layer equations).
\subsubsection{An overview on the classical solution}
In the half plane $(x,y\geq 0)$, by definition of 
\begin{equation}
\eta=\sqrt{\frac{a}{\nu}}~y,
\end{equation}
the velocity vector
\begin{equation}
{\bf {\tilde {u}}}=(\tilde{u},\tilde{v})=(ax\tilde{f}^\prime,-\sqrt{a\nu}\tilde{f}),
\end{equation}
is solenoidal for any arbitrary function $\tilde{f}(\eta)$, defined on half plane $(x,\eta \geq 0)$. In these relations, $\nu$ is the kinematic viscosity, $a$ is an arbitrary constant and $(\cdot)^\prime$ show derivative with respect to $\eta$. Substitution of this velocity vector in the classical Navier--Stokes equations results in
\begin{equation}
\tilde{f}^{\prime\prime\prime}+\tilde{f}{\tilde{f}}^{\prime\prime}-\tilde{f}^{\prime 2}=-1,
\label{stag_e1}
\end{equation} 
for the (up to now arbitrary) function $\tilde{f}(\eta)$.
\begin{figure}[t]
\centering
\includegraphics[width=0.5 \textwidth]{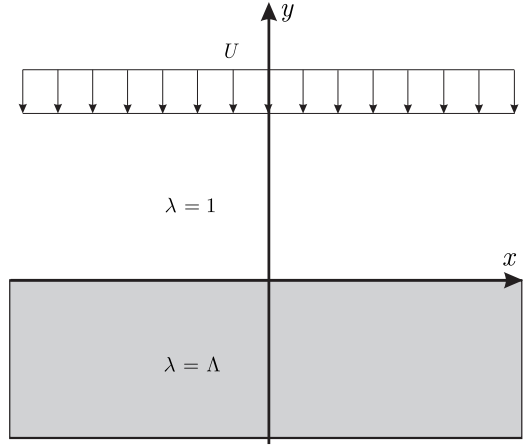}
\caption{Set up of the plane stagnation point flow using the time dilation. The lower half plane has infinity time dilation.} \label{fig9}
\end{figure}
Now, in order to model the viscous stagnation point flow, from all the possibilities for $\tilde{f}$, we are seeking the functions that satisfy the following boundary conditions:
\begin{eqnarray}
&&\tilde{f}(0)=\tilde{f}^\prime(0)=0, \\
&& \tilde{f}(\eta \rightarrow \infty)=1.
\label{stag_e21}
\end{eqnarray}  
The non-linear differential equation (\ref{stag_e1}) with the boundary conditions (\ref{stag_e21}) has not a closed-form solution; although it can be solve numerically, as it is provided in many references.  
\subsubsection{Set up of the new solution}
The plane $(x,y)$ is occupied by an incompressible fluid with the local velocity vector field
\begin{equation}
{\bf {{u}}^*}=({u}^*,{v}^*)=(a^*x{f^*}^\prime,-\sqrt{a^*\nu^*}{f^*}),
\label{Stag_e2}
\end{equation}
in which $a^*$ is a constant with dimension $[a]=T^{-1}$ (and therefore will be scaled by $\lambda$), $\nu^*$ is the molecular kinematic viscosity, and $f^*=f^*(\eta)$ is (for now) an arbitrary smooth function of 
\begin{equation}
\eta^*=\sqrt{\frac{a^*}{\nu^*}}~y,
\end{equation}
and $(\cdot)^\prime$ is derivative with respect to $\eta^*$.

Now, according to Fig. \ref{fig9}, a time dilation is imposed in the lower half plane $\eta\leq 0$ as $\lambda=1+\Lambda {\mathbf H}(\eta)$; and we are seeking a  function $f$ such that the velocity vector 
\begin{equation}
{\bf {u}}=\frac{{\bf u}^*}{\lambda}=(\frac{u^*}{\lambda},\frac{{v}^*}{\lambda})=(\frac{a^*}{\lambda}x{f}^\prime,-\frac{\sqrt{a^*\nu^*}}{\lambda}{f}),
\label{Stag_e21}
\end{equation}
model the plane viscos stagnation point flow, in which $(\cdot)^\prime$ is derivative with respect to $\eta$, where
\begin{equation}
\eta=\sqrt{\frac{a}{\nu}}~y=\sqrt{\frac{a^*}{\nu^*}}~y=\eta^*.
\end{equation}
Note that we defined $a^*=\lambda a$, and like before $\nu^*=\lambda \nu$. Now, these quantities can be used in a similarity solution.
\subsubsection{The similarity equations}
The above velocity vector ${\bf u}$ should satisfy the modified momentum equations, together with a pressure $p=p^*/\lambda^2$. At first, it can easily be verified that the above velocity vector satisfies the continuity equations (it is not shown here.) On the other hand, recalling $\frac{\partial \lambda}{\partial x}=0$, the  modified momentum equations simplify to:
\begin{eqnarray}\label{stag_e2}
({\bf u}^T\cdot \nabla)u &=&-\frac{1}{\varrho^* \lambda^2}\frac{\partial p^*}{\partial x}+\nu\nabla^2{ u}+\underbrace{\frac{1}{\lambda}\left[\nu u \frac{d^2 \lambda}{d y^2}+2\nu \frac{\partial u}{\partial y}\frac{d \lambda}{d y} - uv \frac{d \lambda}{d y}\right]}_{{\tilde F}^\lambda_x},\\ 
({\bf u}^T\cdot \nabla)v&=&-\frac{1}{\varrho^* \lambda^2}\frac{\partial p^*}{\partial y}+\nu\nabla^2{ v}+ \underbrace{\frac{1}{\lambda}\left[\nu v \frac{d^2 \lambda}{d y^2}+2\nu \frac{\partial v}{\partial y}\frac{d \lambda}{d y} - v^2 \frac{d \lambda}{d y}\right]}_{{\tilde F}^\lambda_{y}}.
\end{eqnarray} 
Note that the molecular density is substituted because we want to work with constant quantities; and $p^*$ still is not replaced by $p$ (therefore, the forcing terms are ${\tilde F}^{\lambda}$ not ${F}^{\lambda}$). Now, by substitution from the velocities (\ref{Stag_e21}) 
\begin{eqnarray}\label{stag_e3}
\frac{\lambda^2}{\varrho^* {a^*}^2 x}\frac{\partial p^*}{\partial x}&=&  f^{\prime\prime\prime}+ff^{\prime\prime}-{f^{\prime}}^2 +\underbrace{\frac{\lambda^{\prime\prime}}{\lambda}f^\prime+\frac{\lambda^{\prime}}{\lambda}f^{\prime\prime}+\frac{\lambda^{\prime}}{\lambda}ff^{\prime}}_{{\tilde F}^\lambda_x}    ,\\
-\frac{\lambda^2}{a^* \mu^* }\frac{\partial p^*}{\partial \eta}&=& f^{\prime\prime}+ff^{\prime}+ \underbrace{2\frac{\lambda^\prime}{\lambda}f^\prime+\frac{\lambda^{\prime\prime}}{\lambda}f+\frac{\lambda^{\prime}}{\lambda}f^2}_{{\tilde F}^\lambda_\eta}.
\label{Stage4}
\end{eqnarray} 
The $\eta$--momentum equation is only a function of $\eta$. Therefore, by integration with respect to $\eta$, one can write
\begin{equation}
p^*- p^*_0=a^* \mu^* [\mathcal{K}_1(\eta)+ \mathcal{H}(x)], 
\label{press2} 
\end{equation}
where $\mathcal{K}_1(\eta)$ is a function of $\eta$ will be explained later,  $ \mathcal{H}(x)$ is a function of $x$, $p^*_0$ is the local pressure at the stagnation point, and $p^*$ is the local pressure at each point $(x,\eta)$. This equation and the distribution of pressure will be discussed later. But for now, by differentiation of the above equation with respect to $x$, we have
\begin{equation}
\frac{\partial p^*}{\partial x}=\frac{\partial}{\partial x} \mathcal{H}(x).
\end{equation}
Substitution in the $x$--momentum equation of Eq. (\ref{Stage4}), and taking $\lambda^2$ to the right hand side, one can deduce that the left and right hand sides must be constant, that is
\begin{equation}
 \frac{1}{\lambda^2}\left[ f^{\prime\prime\prime}+ff^{\prime\prime}-{f^{\prime}}^2 +{\tilde F}^\lambda_x \right]={\mathsf C}.
\end{equation}
But this constant is different in the solid and fluid regions. Recalling ${\tilde F}^\lambda_x=0$ for $\eta\neq 0$, the following properties of $f$ is desired:
\begin{eqnarray}
&&{\mathbf{ (1)}~} f^{\prime}=1; \quad f^{\prime\prime\prime}=f^{\prime\prime}=0 \quad {\rm when} \quad \eta\rightarrow \infty; \\
&&{\mathbf{ (2)}~} f^{\prime}=0; \quad f^{\prime\prime\prime}=f^{\prime\prime}=0 \quad {\rm when} \quad \eta\rightarrow -\infty.
\end{eqnarray}
The above requirements will be satisfied if ${\mathsf C}=-1/\lambda^m$, for $m>2$. In the following we have chosen $m=3$, which yields
\begin{equation}
 f^{\prime\prime\prime}+ff^{\prime\prime}-{f^{\prime}}^2 +{\tilde F}^\lambda_x=\frac{-1}{\lambda}.
 \label{stag_eq22}
\end{equation}
in which
\begin{equation}
{\tilde F}^\lambda_x=\frac{\lambda^{\prime\prime}}{\lambda}f^\prime+\frac{\lambda^{\prime}}{\lambda}f^{\prime\prime}+\frac{\lambda^{\prime}}{\lambda}ff^{\prime}.
 \label{stag_eq222}
\end{equation}
This is the similarity equation that must be solved on $\eta\in (-\infty,\infty)$ with suitable boundary conditions (will be discussed later). This equation  is a special case of the Falkner--Skan equation
\begin{equation}
f^{\prime\prime\prime}+ff^{\prime\prime}+\beta[1-(f^\prime)^2]=0,
\end{equation}
with $\beta=1$, and addition of a forcing term ${\tilde F}^\lambda_x$ (and replacing $\beta$ by $\frac{\beta}{\lambda}$). The equation does not have exact solution, and therefore, it should be solved numerically.
\subsubsection{Numerical solution}
Traditionally, such two-point boundary value problems are solved using the shooting method, in which the equation is integrated iteratively form $\eta=0$ to $\eta=\eta_\infty\rightarrow \infty$ with a trial value for $f^{\prime\prime}$. However, in Eq. (\ref{stag_eq22}) there are penalization terms that work properly when we solve the problem as a boundary value problem not an initial value problem. Moreover, although the equation is third order, the no-slip condition is implementing by the penalization terms, therefore, we need just two boundary conditions. Because of these issues, we have to solve Eq. (\ref{stag_eq22})--(\ref{stag_eq222}) as a boundary value problem.

To this end, we adopted the method of Asaithambi [1] here (originally proposed for solution of the Falkner--Skan equation). The method contains two main steps. At first, by definition of a length scale $\eta_\infty$, the equation is transformed to a new coordinate $\xi=\eta/\eta_\infty$ as
\begin{equation}
\frac{1}{\eta_\infty^3} f^{\prime\prime\prime}+\frac{1}{\eta_\infty^2}ff^{\prime\prime}-\frac{1}{\eta_\infty^2}{f^{\prime}}^2 +\frac{1}{\eta_\infty}\frac{\lambda^{\prime\prime}}{\lambda}f^\prime+\frac{1}{\eta_\infty^2}\frac{\lambda^{\prime}}{\lambda}f^{\prime\prime}+\frac{1}{\eta_\infty}\frac{\lambda^{\prime}}{\lambda}ff^{\prime}=\frac{-1}{\lambda},
\end{equation}
where primes are derivatives with respect to $\xi$.

The next step is to define an auxiliary variable
\begin{equation}
u=\frac{1}{\eta_\infty}\frac{d f}{d\xi},
\label{auxiliary1}
\end{equation}
and substitution to the above equation, which yields:
\begin{equation}
u^{\prime\prime}+\eta_\infty (f+\frac{\lambda^\prime}{\lambda})u^\prime+\eta_\infty^2 \left[ \frac{1}{\lambda} +(\frac{\lambda^\prime}{\lambda}f+\frac{\lambda^{\prime\prime}}{\lambda})u-u^2 \right]=0.
\label{ODE1}
\end{equation}
Now, by second--order central finite-difference discretization of the equation on a uniform grid $\xi_j$ that $j=0,\cdots,(N-1)$ one obtains
\begin{eqnarray}
{\rm G}_j({\bf u},{\bf f})&=&\left[1-\frac{1}{2}\eta_\infty(f_j+\frac{\lambda^{\prime}}{\lambda})\Delta\xi\right]u_{j-1}- \left[2-\eta_\infty^2(\frac{\lambda^{\prime}}{\lambda}f_j+\frac{\lambda^{\prime\prime}}{\lambda}-u_j)\Delta\xi^2\right]u_j \nonumber \\
&& +\left[1+\frac{1}{2}\eta_\infty(f_j+\frac{\lambda^{\prime}}{\lambda})\Delta\xi\right]u_{j+1}+\frac{\eta_\infty^2\Delta \xi^2}{\lambda}=0,
\end{eqnarray} 
in which ${\bf u}=[u_0\cdots~u_{N-2}]^T$ and ${\bf f}=[f_0 \cdots f_{N-2}]^T$. On the other hand, the auxiliary equation Eq. (\ref{auxiliary1}) is discretized as
\begin{equation}
f_j=f_{j-1}+\frac{1}{2}\eta_\infty(u_j+u_{j-1})\Delta \xi.
\label{auxiliary2}
\end{equation}
The solution methodology is based on iterative Newton solution of the non-linear system 
\begin{equation}
{\rm{\bf G}}({\bf u},{\bf f})=0
\label{nonlinear}
\end{equation}
with the auxiliary equation (\ref{auxiliary2}). \\
With this regards, linearization of Eq. (\ref{nonlinear}) yields
\begin{equation}
{\rm {\bf J}}({\rm {\bf u}}^k) \Delta {\rm {\bf u}}^k =-{\rm{\bf G}}({\bf u}^k,{\bf f}^k), \label{Newton}
\end{equation}
in which the $j$-th row of the (tri-diagonal) Jacobean matrix ${\rm {\bf J}}$ has the entities
\begin{eqnarray}
{\rm {\bf J}}_{j-1}&=&1-\frac{\eta_\infty}{2}\Delta \xi  \left(f_j+\frac{\lambda^\prime}{\lambda}\right), \nonumber \\
{\rm {\bf J}}_{j}&=& -2+\eta_\infty^2\Delta\xi^2\left[\frac{\lambda^\prime}{\lambda}f_j+\frac{\lambda^{\prime\prime}}{\lambda}-2u_j \right] ,\\
{\rm {\bf J}}_{j+1}&=&1+\frac{\eta_\infty}{2}\Delta \xi  \left(f_j+\frac{\lambda^\prime}{\lambda}\right). \nonumber
\end{eqnarray} 
Solution of Eq. (\ref{Newton}), results in ${\bf u}^{k+1}$ via
\begin{equation}
{\mathbf u}^{k+1}={\mathbf u}^{k}+\Delta {\mathbf u}^{k}, \label{Newton2}
\end{equation}
which complete the needed equations for the iterative solution of Eq. (\ref{stag_eq22})--(\ref{stag_eq222}).\\ 
The algorithm of Asaithambi [1] is designed so that $\eta_\infty$ be found as a part of solution. In the present work, we assume that $\eta_\infty=5$ has already known (based on the last studies of the classical stagnation point flow). Therefore, our algorithm has one loop (in contrast to the original algorithm of Asaithambi that has two loops):
\begin{itemize}
\item[{\bf (0)}] Begin from $\eta_\infty=5$ and ${\bf u}={\bf f}=0$; on a $N$-point uniform grid $\xi_j$, $j=0,\ldots,N-1$.
\item[{\bf (1)}] Equation (\ref{Newton}) is solved (by using a classical Thomas algorithm) with the boundary conditions $\Delta u_0=\Delta u_{N-1}=0$
\item[{\bf (2)}] Velocity $\mathbf u$ is updated from Eq. (\ref{Newton2}), with the boundary conditions $u_0=0,u_{N-1}=1$. 
\item[{\bf (3)}] $\mathbf f$ is updated from Eq. (\ref{Newton2}), and the boundary conditions $f_0=0$.
\item[{\bf (4)}] Repeat steps {\bf (1)} to {\bf (3)} until convergence.
\end{itemize}
\begin{figure}[t]
\centering
\includegraphics[width=1 \textwidth]{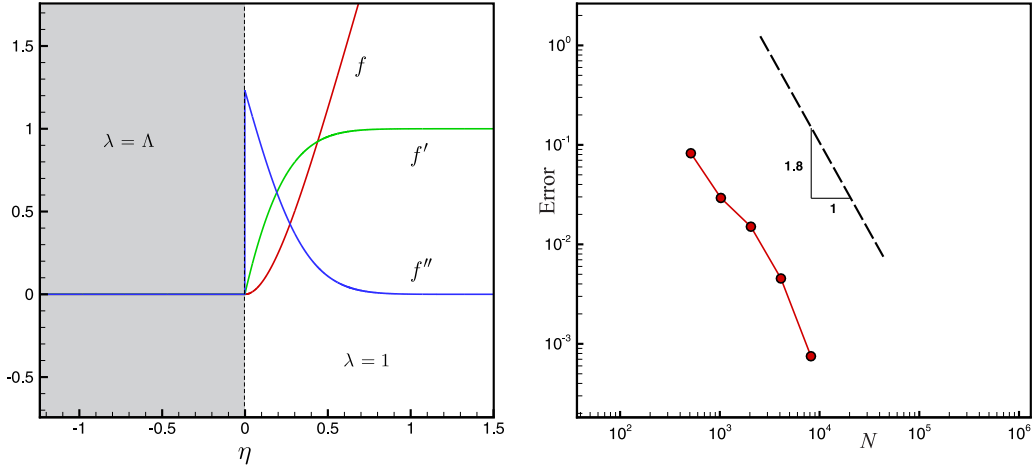}
\caption{The results of numerical solution of equations (\ref{stag_eq22})--(\ref{stag_eq222}). Left:  $f,f^\prime$ and $f^{\prime\prime}$ are shown versus $\eta$. The results are in very good agreement with the classical solutions. Right: Rate of decaying of the errors is studied. The errors are defined as the difference between $f_N^{\prime\prime}(0)$ calculated on the $N$-point grid and the reference value $f_{N_{\rm max}}^{\prime\prime}(0)=1.23260377$. An ${\mathcal O}(1.8)$ is found in the mean.} \label{stagfig2}
\end{figure}
The above algorithm was applied on the uniform grids with different resolutions from $N_{\rm min}=512+1$ to $N_{\rm max}=16384+1$, and the convergence rate was obtained. The Heaviside function ${\mathbf H}(\eta)$ is constructed by a scaled error function ${\rm erf}(\mathit{S}\eta)$, where $\mathit{S}$ is a scale factor which controls the sharpness of the Heaviside function. The derivatives $\lambda^\prime$ and $\lambda^{\prime\prime}$ are approximated by second order central finite--differencing of the Heaviside function.  On all grids the maximum sharpness $\Delta\sim 1/N$ were chosen and in all runs, $\Lambda=10^{30}$.

The results are shown in Fig. \ref{stagfig2}. In the left panel, $f,f^\prime$ and $f^{\prime\prime}$ are shown on $-1.5\leq\eta\leq 1.5$. As one can see, the no-slip condition is captured correctly. Moreover, note that all the three quantities are vanished in the solid region $\eta\leq 0$. 

In the right panel, the convergence rate of the numerical solution is studied. Since there is not an exact solution for this flow, we defined our error based on the $f^{\prime\prime}(\eta=0)$. This quantity, which is in relation with the wall stresses, is the free parameter in the conventional shooting method, and should be determined correctly in order to have an accurate solution. As the reference value we used $f_{\rm Ref.}^{\prime\prime}(0)=1.23260377$ which is obtained on a $N_{\rm max}$-point grid using the Asaithambi method for the classical stagnation point flow. The errors are calculated in comparison with this reference value. As one can see, the rate of decaying of errors is of ${\mathcal O}(1.8)$ in the mean. \\

\noindent
{\bf Pressure distribution} \\
As it was mentioned earlier, the $\eta$--momentum of Eq. (\ref{Stage4}) can be integrated directly with respect to $\eta$ that gives Eq. (\ref{press2}). In this equation, it is not so difficult to show that $ \mathcal{H}(x)\sim x^2$, and therefore, $ \mathcal{H}(0)=0$. Consequently, on the stagnation line, Eq. (\ref{press2})  can be written as
\begin{equation}
p^*- p^*_0=a^* \mu^* \mathcal{K}_1(\eta), 
\label{press3} 
\end{equation}
where
\begin{equation}
\mathcal{K}_1(\eta)=\int_{0}^{\eta}\frac{1}{\lambda^2} \left( f^{\prime\prime}+ff^{\prime}+ {\tilde F}^\lambda_\eta \right) d \eta. 
\label{press4} 
\end{equation}
At first, by a looking at $f(\eta)$ and $f^\prime(\eta)$ that we found, one can see that ${\tilde F}^\lambda_\eta=0$; and the above integral can be written as 
\begin{equation}
\mathcal{K}_1(\eta)=\int_{0}^{\eta}\frac{1}{\lambda^2} \frac{d}{d\eta} \left( f^{\prime}+\frac{1}{2} f^2 \right) d \eta. 
\label{press5} 
\end{equation}
We treat the fluid and solid regions in separation:
\begin{itemize}
\item[{\bf (1)}] For the fluid region $\eta\geq 0$, we have $\lambda=1$, and
\begin{equation}
p^*-p_0^*=p-p_0=a^*\mu^*\left( f^{\prime}+\frac{1}{2} f^2 \right), 
\label{press6} 
\end{equation}
which is in agreement with the classical plane stagnation point flow solutions.
\item[{\bf (2)}] For the solid region $\eta< 0$, we have $\lambda\rightarrow \infty$, and
\begin{equation}
\frac{d p^*}{d\eta}=0 \Longrightarrow  p^*=p_0^*=p_0. 
\label{press7} 
\end{equation}
Now, since $p^*=\lambda^2 p$, we can write:
\begin{equation}
p=\frac{p_0}{\lambda^2}\rightarrow 0. 
\label{press8} 
\end{equation}
\end{itemize}
Note that although in the solid region $p^*$ remains $p_0$, but jump in $p$ is possible because of the $\lambda^2$ coefficient. In fact $p\rightarrow 0$ suddenly at the solid boundary.\\  
\begin{figure}[t]
\centering
\includegraphics[width=0.65 \textwidth]{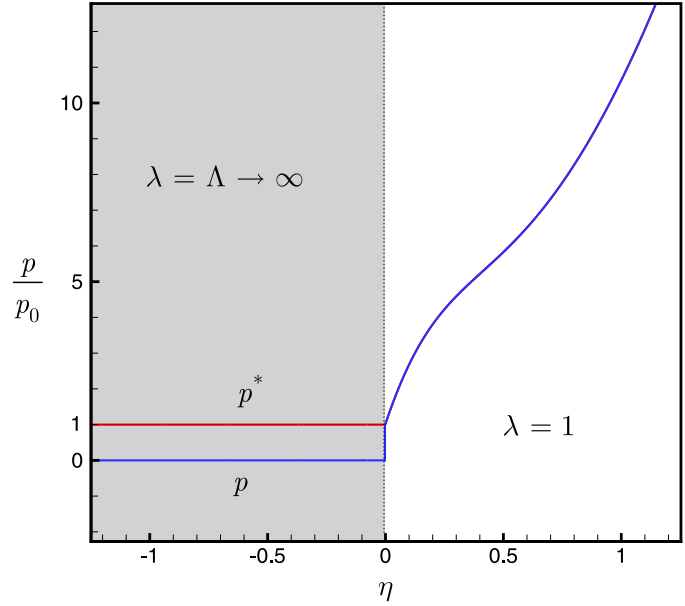}
\caption{Pressure distribution in fluid and solid regions for the plane stagnation point flow. In the solid region $p^*$ remains equal to $p_0$, while $p$ has a jump to zero.} \label{stagfig3}
\end{figure}

In Fig. \ref{stagfig3} distribution of $p/p_0$ and $p^*/p_0$ along the stagnation line are illustrated  for $a^*\mu^*=1$. In this figure $f$ and $f^\prime$ are substituted from our numerical solution. 
\subsection{Stokes flow over a sphere}
The modified equations for the Stokes flow are obtained, in the spherical coordinate, and the Stokes flow over a sphere is re-solved using them. In contrast to the classical Stokes solution, the new velocities and pressure vanish inside the sphere. 
\subsubsection{An overview on the classical solution}
\begin{figure}[t]
\centering
\includegraphics[width=0.55 \textwidth]{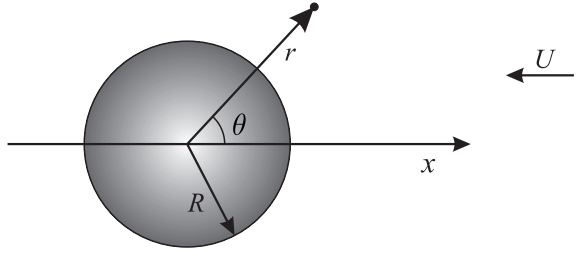}
\caption{Stokes flow over a sphere. A solid sphere of radius $R$ is placed in a uniform velocity field $(-U{{\bf \hat{e}}}_x)$, such that ${\rm Re}=\frac{\varrho^*UR}{\mu^*}\rightarrow 0$.} \label{fig10}
\end{figure}

According to Fig. \ref{fig10}, a solid sphere of radius $R$ is placed in a uniform velocity field $(-U{{\bf \hat{e}}}_x)$, of a Newtonian fluid with the molecular viscosity $\mu^*$, such that the Reynolds number ${\rm Re}=\frac{\varrho^*UR}{\mu^*}\rightarrow 0$. Therefore, the flow field is governed approximately by the Stokes equations. Recalling $\partial/\partial \phi=0$, these equations can be written in the spherical coordinates as
\begin{eqnarray}
&&\frac{1}{\mu^*}\frac{\partial p^*}{\partial r} =\nabla^2 u^*_r-\frac{2u^*_r}{r^2}-\frac{2}{r^2\sin \theta}\frac{\partial}{\partial \theta} (u^*_\theta \sin{\theta}),\label{sphere0} \\
&&\frac{1}{\mu^*}\frac{1}{r}\frac{\partial p^*}{\partial \theta}  =\nabla^2 u^*_\theta-\frac{u^*_\theta}{r^2\sin^2\theta}+\frac{2}{r^2}\frac{\partial u^*_r}{\partial \theta},\label{sphere1}\\
&&\frac{1}{r^2}\frac{\partial}{\partial r}(r^2 u^*_r)+ \frac{1}{r\sin \theta}\frac{\partial}{\partial \theta}(u^*_\theta \sin\theta)=0,
\label{sphere2}
\end{eqnarray} 
in which $\nabla^2$ is defined as
\begin{equation}
\nabla^2=\frac{1}{r^2}\frac{\partial}{\partial r}\left( r^2 \frac{\partial}{\partial r} \right)+\frac{1}{r^2 \sin\theta}\frac{\partial}{\partial \theta}\left( \sin \theta \frac{\partial}{\partial \theta} \right). \label{nabla2}
\end{equation}
Using the separation of variables technique, Stokes admirably found  a solenoidal velocity vector  ${\bf u}^{\rm St}=(u^{\rm St}_r,u^{\rm St}_\theta)$ that satisfies the no-slip condition at the sphere surface
\begin{eqnarray}
u^{\rm St}_r&=&U\left[ -\frac{1}{2}(\frac{R}{r})^3+\frac{3}{2}(\frac{R}{r})-1\right] \cos\theta, \label{St1}\\
u^{\rm St}_\theta&=&U\left[ -\frac{1}{4}(\frac{R}{r})^3-\frac{3}{4}(\frac{R}{r})+1\right] \sin\theta. \label{St2}
\end{eqnarray}
By substituting these velocities in the momentum equations and integrating, one obtains the pressure distribution 
\begin{equation}
p^{\rm St}-p_\infty=\frac{3}{2}(\frac{R}{r})^2\cos\theta.
\end{equation} 
The Stokes solution $({\bf u}^{\rm St},p^{\rm St})$ models the fluid flow outside the solid sphere very satisfactorily such that the resulting stresses and the drag force are in very good agreements with experiments. Not only  ${\bf u}^{\rm St}$ satisfies the no-slip condition, but also, surprisingly, it satisfies partially  the no-diffusion condition. In fact, by a looking at the radial velocity (\ref{St1}), one can see that
\begin{equation}
\nabla {\bf u}^{\rm St}\cdot {\bf \hat{n}}|_{r=R}=\nabla_r{\bf u}^{\rm St}|_{r=R} =\frac{\partial}{\partial r}u^{\rm St}_r(R,\theta)=0.
\end{equation}
However, the no-diffusion condition is not satisfying completely, because (by looking at Eq. (\ref{St2})), one can see
\begin{equation}
\frac{\partial}{\partial r}u^{\rm St}_\theta(R,\theta) \neq 0.
\end{equation}
Moreover, the velocity components are not vanishing inside the sphere. In fact,$({\bf u}^{\rm St},p^{\rm St})\rightarrow (-\infty,+\infty)$ at the center of the sphere.
\subsubsection*{The new solution}
In a uniform velocity field $(-U{{\bf \hat{e}}}_x)$ of a Newtonian fluid with the molecular viscosity $\mu^*$, the time dilation $\lambda=1+\Lambda H(R)$ is implemented that $R$ is small enough such that ${\rm Re}=\frac{\varrho^*UR}{\mu^*}\rightarrow 0$. By substitution of $({\bf u}^*,p^*)=(\lambda{\bf u},\lambda p)$ in the Stokes equations (\ref{sphere0})--(\ref{sphere2}), one can find the modified Stokes equations as
\begin{eqnarray}
&&\frac{1}{\mu}\frac{\partial p}{\partial r} =\nabla^2 u_r-\frac{2u_r}{r^2}-\frac{2}{r^2\sin \theta}\frac{\partial}{\partial \theta} (u_\theta \sin{\theta})+F^{\lambda}_r,\label{sphere6} \\
&&\frac{1}{\mu}\frac{1}{r}\frac{\partial p}{\partial \theta}  =\nabla^2 u_\theta-\frac{u_\theta}{r^2\sin^2\theta}+\frac{2}{r^2}\frac{\partial u_r}{\partial \theta}+F^{\lambda}_\theta,\label{sphere7}\\
&&\frac{1}{r^2}\frac{\partial}{\partial r}(r^2 u_r)+ \frac{1}{r\sin \theta}\frac{\partial}{\partial \theta}(u_\theta \sin\theta)=0, \label{sphere8}
\label{sphere3}
\end{eqnarray} 
in which
\begin{eqnarray}
&&F^{\lambda}_r=\left( \frac{\lambda^{\prime\prime}}{\lambda}+\frac{2}{r} \frac{\lambda^{\prime}}{\lambda}\right)u_r+2\frac{\lambda^{\prime}}{\lambda}\frac{\partial u_r}{\partial r}-\frac{2}{\mu}\frac{\lambda^{\prime}}{\lambda}p, \label{sphere4}\\
&&F^{\lambda}_\theta=\left( \frac{\lambda^{\prime\prime}}{\lambda}+\frac{2}{r} \frac{\lambda^{\prime}}{\lambda}\right)u_\theta+2\frac{\lambda^{\prime}}{\lambda}\frac{\partial u_\theta}{\partial r}, \label{sphere5}
\end{eqnarray}
where $\nabla^2$ is defined in (\ref{nabla2}), the primes show the derivatives with respect to $r$,  and $\mu=\mu^*/\lambda$.\\
It should be noted that the no-advection condition is absent in the modified Stokes equations (\ref{sphere6})--(\ref{sphere5}), because the advection terms are ignored in the original Stokes equations.
\begin{figure}[t]
\centering
\includegraphics[width=0.65 \textwidth]{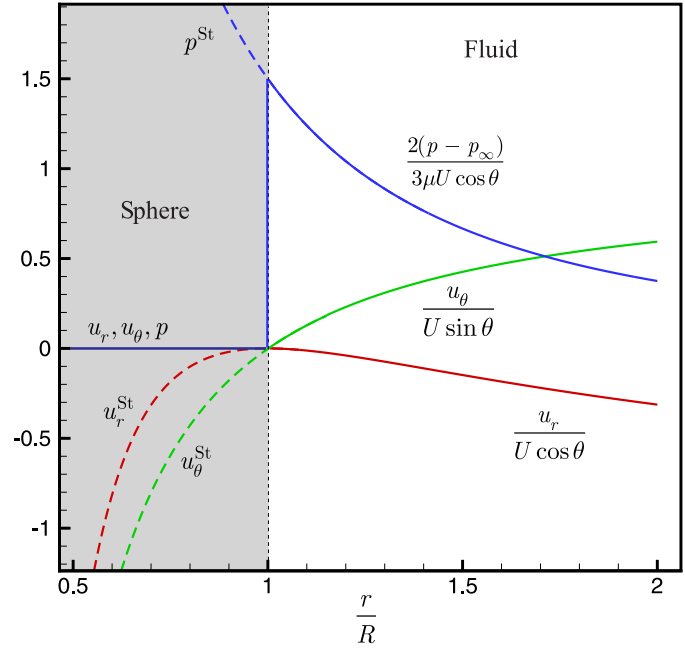}
\caption{Comparison of the classical and new solutions of the Stokes flow over a sphere. The classical velocities $(u_r^{\rm St},u_\theta^{\rm St})$ and pressure $p^{\rm St}$ are not vanished inside the sphere. In fact $({\bf u}^{\rm St},p^{\rm St})\rightarrow (-\infty,+\infty)$ at the center of sphere. In the new solution, all the quantities go to zero inside the sphere. Partiulary, jump of the pressue  at the bounday is observable.} \label{fig11}
\end{figure}

It is not difficult to verify that the Stokes solution $(u_r^{\rm St},u_\theta^{\rm St},p^{\rm St})$ does not satisfy the modified Stokes equations (\ref{sphere6})--(\ref{sphere5}). On the other hand, we define the exact solution as  
\begin{equation}
({\bf u},p)=({\bf u}^{\rm St}/\lambda,p^{\rm St}/\lambda^2). \label{exact2}
\end{equation}
Particularly, for the radial pressure distribution, by substitution of ${\bf u}^{\rm St}$ in the right hand side of the radial momentum equation (\ref{sphere6}) and (\ref{sphere4}) we have
\begin{equation}
\lambda \left[ \frac{1}{\mu}\frac{\partial p}{\partial r}+\frac{2}{\mu}\lambda^\prime p \right]=\nabla^2 u^{\rm St}_r-\frac{2u^{\rm St}_r}{r^2}-\frac{2}{r^2\sin \theta}\frac{\partial}{\partial \theta} (u^{\rm St}_\theta \sin{\theta}),
\end{equation}
which means
\begin{equation}
\frac{1}{\mu^*}\frac{\partial}{\partial r}(\lambda^2p)=\nabla^2 u^{\rm St}_r-\frac{2u^{\rm St}_r}{r^2}-\frac{2}{r^2\sin \theta}\frac{\partial}{\partial \theta} (u^{\rm St}_\theta \sin{\theta});
\end{equation}
therefore
\begin{equation}
\lambda^2p=p^{\rm St},
\end{equation}
which means 
\begin{equation}
p=\frac{p^{\rm St}}{\lambda^2}\rightarrow 0 \quad \quad {\mathrm{for}} \quad r<R.
\end{equation}
The radial distribution of the velocities and pressure are illustrated in Fig.  \ref{fig11}. As one can see, the new solution goes to zero inside the sphere.\\

In summary, we showed that the solution of the modified Stokes equations is equal to the Stokes solution in the fluid side, and it goes to zero in the sphere.  
\section{Conclusions}
An approach for exact imposition of the immersed solid boundaries on the incompressible Navier--Stokes equations has been proposed. The method consists of modification of the Navier--Stokes equations so that involve arbitrary time dilations. The modified Navier--Stokes equations have some penalization terms in the right hand side, each one discards one mechanism of coupling of inside and outside of the solid body, that is, the advection, diffusion, and pressure coupling. The method is applied on three classical exact solutions of the Navier--Stokes equation, that is, the Stokes first problem, the plane stagnation point flow, and the Stokes flow over a sphere; and it has been shown that the exact solutions are obtainable in the presence of solid rigid bodies.  

%
\end{document}